\documentclass[11pt]{article}
\pdfoutput=1
\usepackage{amsmath}
\usepackage{amssymb}
\usepackage{graphicx,bbm,mathrsfs}
\usepackage{nicefrac}
\usepackage{bbm}
\usepackage{geometry}
\usepackage{jheppub}

\usepackage{dcolumn}   
\usepackage{bm}        
\usepackage{graphicx,mathrsfs}
\usepackage{nicefrac}
\usepackage{multirow}
\usepackage{color}

\newcommand{\be}{\begin{equation}}
\newcommand{\ee}{\end{equation}}
\newcommand{\bea}{\begin{eqnarray}}
\newcommand{\eea}{\end{eqnarray}}

\newcommand{\eq}[1]{\begin{align}#1\end{align}}
\newcommand{\Ref}[1]{Ref.~\cite{#1}}
\newcommand{\Refs}[1]{Refs.~\cite{#1}}

\newcommand{\Eq}[1]{Eq.~\eqref{#1}}
\newcommand{\Eqs}[2]{Eqs.~\eqref{#1} and \eqref{#2}}
\newcommand{\Eqss}[3]{Eqs.~\eqref{#1}, \eqref{#2} and \eqref{#3}}
\newcommand{\Sec}[1]{Sec.~\ref{#1}}
\newcommand{\Secs}[2]{Secs.~\ref{#1} and \ref{#2}}

\newcommand{\mPl}{M_{\rm Pl}}
\newcommand{\mPlr}{M_{\rm Pl}}
\newcommand{\GN}{G_{\rm N}}
\newcommand{\Lgrav}{\Lambda_\mathrm{grav}}
\newcommand{\MBH}{M_\mathrm{BH}}
\newcommand{\RBH}{R_\mathrm{BH}}
\newcommand{\TBH}{T_\mathrm{BH}}
\newcommand{\MBHm}{M_*}

\newcommand{\nng}{n_{\not G}}
\newcommand{\cng}{c_{\not G}}

\usepackage{wasysym}

\usepackage{titlesec}

\titleformat*{\section}{\Large\bfseries}
\titleformat*{\subsection}{\large\bfseries}
\titleformat*{\subsubsection}{\large\bfseries}
\titleformat*{\paragraph}{\large\bfseries}
\titleformat*{\subparagraph}{\large\bfseries}

\makeatletter
\newcommand*{\prodsym}{%
  \DOTSB
  \mathop{
    \mathchoice
      {\rlap{\kern.3em\rotatebox[origin=c]{-90}{}}{\prod}}
      {\vcenter{\rlap{\kern.2em\rotatebox[origin=c]{-90}{}}}{\prod}}
      {\sum}{\sum}
  }\slimits@
}
\makeatother

\begin{document}
	\hfill CALT-TH-2019-032

\vspace*{1.2cm}

\begin{center}

\thispagestyle{empty}
{\LARGE \bf
Approximate Symmetries and Gravity
 }\\[10mm]

\renewcommand{\thefootnote}{\fnsymbol{footnote}}

{\large  Sylvain~Fichet$^{a,b}$\footnote{sfichet@caltech.edu },\, 
Prashant~Saraswat$^a$\footnote{saraswat@caltech.edu }}\\[10mm]

\end{center} 
\noindent
\quad\quad\quad\textit{$^a$ Walter Burke Institute for Theoretical Physics, California Institute of Technology,}

\noindent \quad\quad\quad \textit{Pasadena, CA
91125, California, USA} \\

\noindent
\quad\quad\quad \textit{$^b$ ICTP South American Institute for Fundamental Research  \& IFT-UNESP,}

\noindent \quad\quad\quad \textit{R. Dr. Bento Teobaldo Ferraz 271, S\~ao Paulo, Brazil
}
2019-032
\addtocounter{footnote}{-2}

\vspace*{12mm}

\begin{center}
{  \bf  Abstract }
\end{center}

There are strong reasons to believe that global symmetries of quantum theories cannot  be exact in the presence of gravity.  While this has been argued at the qualitative level, establishing a quantitative statement is more challenging. In this work we take new steps towards quantifying symmetry violation in EFTs with gravity.    First,  we evaluate global charge violation by microscopic black holes present in a thermal system, which represents an irreducible, universal effect at finite temperature.    Second, based on general QFT considerations,  we propose that local symmetry-violating processes should be faster than black hole-induced processes at any sub-Planckian temperature. 
    
    Such a proposal can be seen as part of the ``swampland" program to constrain EFTs emerging from quantum gravity. Considering an EFT perspective, we formulate a conjecture which requires the existence of operators violating global symmetry and places quantitative bounds on them.
We study the interplay of our conjecture with emergent symmetries in QFT. In models where gauged $U(1)$'s enforce accidental symmetries, we find that constraints from the Weak Gravity Conjecture can ensure that our conjecture is satisfied. We also study the consistency of the conjecture with QFT models of emergent symmetries such as extradimensional localization, the Froggatt-Nielsen mechanism, and the clockwork mechanism.

\newpage
\enlargethispage{\baselineskip}
\tableofcontents
\newpage

\section{Introduction}

\label{se:intro}

Symmetries have long been an important conceptual tool for understanding quantum field theories (QFTs) and quantum mechanics in general. There is a crucial distinction between gauge (or local) symmetries and global symmetries: the former describe mathematical redundancies in the description of the Hilbert space of a theory, while the latter genuinely act as unitary transformations on quantum states.\,\footnote{It is therefore more appropriate to speak of ``gauge redundancy'' than ``gauge symmetry," as is often noted. Accordingly, we will often simply use the word ``symmetries'' to refer to specifically global symmetries in this work.} While gauge symmetry is a necessary ingredient for defining certain Lorentz-invariant unitary QFTs, global symmetries are instead additional postulates regarding the form of a theory, which give useful constraints on the dynamics. For example, a continuous global symmetry implies an exactly conserved current and/or an exactly massless Goldstone boson (if non-linearly realized, \textit{i.e.} spontaneously broken).      

Even in the absence of exact global symmetries, the concept of approximate symmetries is an extremely useful tool for studying QFTs. For example, the Standard Model features approximate chiral symmetries (broken by small Yukawa couplings), which contribute to the understanding of the pion mass and suppression of flavor-changing neutral currents. Implicit in the application of approximate symmetries is the idea that one can consider continuously taking a limit in which the symmetry is exact. Corrections to exact conservation laws are then proportional to some small parameter.

In light of the usefulness of global symmetries, it is somewhat surprising that there are strong arguments (reviewed below) that exact continuous global symmetries cannot exist in quantum theories with gravity (see \textit{e.g.} \Ref{Banks:2010zn}). With this in mind, one should not view models with approximate symmetries as continuous deformations of some theory with exact symmetries. Approximate symmetries must be emergent in effective field theory (EFT) descriptions of quantum gravity (QG) theories, suggesting that any small parameters facilitating the approximate symmetry should ultimately have dynamical origins instead of being regarded as free parameters of the theory. 

An argument against global symmetries in theories of quantum gravity can be made without reference to specific UV completions such as string theory; one need only consider semiclassical properties of black holes which can be understood from low-energy physics. Black holes are thermodynamic objects with a finite Bekenstein-Hawking entropy $S = A/(4 G_N)$ indicating the number of microphysical quantum states represented by a classical black hole solution. If a continuous global symmetry is exactly conserved, then this entropy should be able to represent all possible global charges that could be carried by a black hole. However, since global charges are not associated with a long-range gauge field, a black hole could carry arbitrarily large global charge without any modification to its classical description, in contradiction with the finiteness of the Bekenstein-Hawking entropy. The same argument suggests that even for a gauge symmetry one cannot take the limit in which the gauge coupling constant $g$ goes to zero so that the gauge field decouples, helping to motivate the ``Weak Gravity Conjecture'' (WGC)~\cite{ArkaniHamed:2006dz}. In an independent line of reasoning, \Refs{Harlow:2018tng, Harlow:2018jwu} showed that the AdS/CFT correspondence does not allow exact global symmetries in the bulk.

Despite the expectation that quantum gravity violates global symmetries, it seems difficult to derive symmetry violating effects directly from semi-classical gravity considerations. One approach is the study of wormholes, which can arise for instance as solutions of Euclidean gravity in the presence of an axion~\cite{Giddings:1987cg}. If included in the path integral they generate an effective potential for the axion violating its shift symmetry~\cite{Rey:1989mg, Kallosh:1995hi, Alonso:2017avz}. More broadly however they are expected to induce all local operators in the theory~\cite{Coleman:1988cy, Giddings:1988cx}, potentially including explicit symmetry-violating operators~\cite{Abbott:1989jw} and the cosmological constant~\cite{Coleman:1988tj}. However, the interpretation and consistency of these Euclidean wormhole effects is a subject of debate (see \Ref{Hebecker:2018ofv} for a recent review). \Refs{ArkaniHamed:2006dz, Rudelius:2015xta, Brown:2015iha, Heidenreich:2015wga} proposed a lower bound on axion potentials by analogy with the Weak Gravity Conjecture; however this does not directly constrain the case of linearly realized global symmetries.\footnote{In \Ref{Hebecker:2019vyf} a slightly different bound on axion masses was proposed, which in certain scenarios could also imply lower bounds on symmetry-violating fermionic interactions.}

In the absence of any rigorous derivation starting from quantum gravity, we can consider the principle that there are no exact global symmetries as an element of the ``swampland'' program~\cite{Vafa:2005ui, Ooguri:2006in}, which aims to find constraints on the form of EFTs arising as low-energy descriptions of quantum gravity (see~\cite{Palti:2019pca} for a recent review of the swampland program). While the arguments against exact global symmetries have long been known, there has been relatively little progress in constraining \emph{approximate} symmetries and delineating them from the (inconsistent) case of exact symmetries. This situation can be contrasted to the quantitative understanding of the obstacles to taking gauge couplings to zero, which are expressed in the various forms of the Weak Gravity Conjecture (\textit{e.g.} \Refs{ArkaniHamed:2006dz, Cheung:2014vva, Heidenreich:2015nta, Heidenreich:2016aqi, Andriolo:2018lvp}).  

Any potential quantitative constraint on the exactness of global symmetries must be consistent with the fact that approximate symmetries can quite easily emerge in the low-energy limit of a more complete theory. This is already familiar from the Standard Model. Baryon and lepton number emerge as ``accidental'' symmetries of the SM since the gauge symmetry forbids any relevant or marginal operators violating them. Any such operator is necessarily dimension 5 or greater and its effects are greatly suppressed at energies below the SM cutoff. The approximate chiral and flavor symmetries of the SM could also potentially be realized as emergent symmetries. For example, models of flavor based on extra-dimensional localization~\cite{Grossman:1999ra} often assume no approximate global symmetries at all in the higher-dimensional theory, yet in the 4D limit produce exponentially small violations of global symmetries  in the form of small Yukawa couplings. Froggatt-Nielsen models~\cite{Froggatt:1978nt} similarly can produce exponentially small couplings in the low-energy EFT from the appearance of a vev in high-dimension operators, without necessarily invoking a global symmetry. All of these examples suggest that any swampland constraint on global symmetries should allow EFTs with a low cutoff to have exponentially small violations of symmetry.

The following question then naturally arises: Just how exact can a global symmetry of an EFT be in the presence of gravity? In this work we take two steps towards quantitatively constraining this. First, we compute the rate $\Gamma^{\not G}_{\rm BH}$ at which (approximate) global charges are destroyed in a thermal system due to the effects of black holes, which exist in a thermal bath with a Boltzmann-suppressed number density. This gives an irreducible rate for the violation of global charge in any theory with gravity, as a function of temperature. Second, we conjecture that in any QG theory there are sub-Planckian processes which can be described in an EFT framework and which violate global charge in a thermal system at a rate $\Gamma^{\not G}_{\rm EFT}$ which is at least as fast as the effect from black holes. We provide some qualitative motivations for this proposal, which we will refer to as the ``Swampland Symmetry Conjecture'' (SSC) (\Sec{se:swamp}), and check its validity in settings where approximate global symmetries emerge at low energies. For example, we find that the Weak Gravity Conjecture enforces the Swampland Symmetry Conjecture in models where accidental global symmetries emerge due to constraints from gauge symmetry (\Sec{se:WGC}). We also consider various QFT mechanisms for suppressing violations of global symmetry at low energies (\Sec{se:Models}), and find that if these are implemented at sub-Planckian energies then the resulting low-energy EFTs manifestly satisfy the SSC as well.

\section{Black holes, effective theory and global symmetries }
\label{se:general}

Though the high-energy nature of quantum gravity is unknown, at sub-Planckian energies general relativity can be treated as an effective quantum field theory (EFT). Graviton amplitudes are just described by Feynman diagrams of arbitrary complexity, which is well defined and in a sense  already is a theory of quantum gravity.

General relativity also has black hole solutions, which occur in the real world. Although macroscopic black holes remain mysterious objects in some ways, for instance from the viewpoint of information, other aspects of their physics are well established. Their formation from matter states is described using critical density considerations, and their evaporation is induced by Hawking radiation.

However, the graviton EFT and the theory of black holes are not  unified. Although some semiclassical estimates are possible, there is no precise description of a quantum black hole in either theory. Some exciting progress in the field theoretical description of black holes have been made recently, see e.g. \cite{Dvali:2014ila,Dvali:2016ovn}, but a full understanding of black holes within a QFT framework remains out of reach thus far. It is in principle possible that future progress in the treatment of the graviton QFT could provide a description of quantum black holes, which would then constitute a full theory of quantum gravity in the sub-Planckian regime.

\begin{figure}
\center
\includegraphics[width=10 cm,trim={3.cm 4cm 7.cm 3cm},clip]{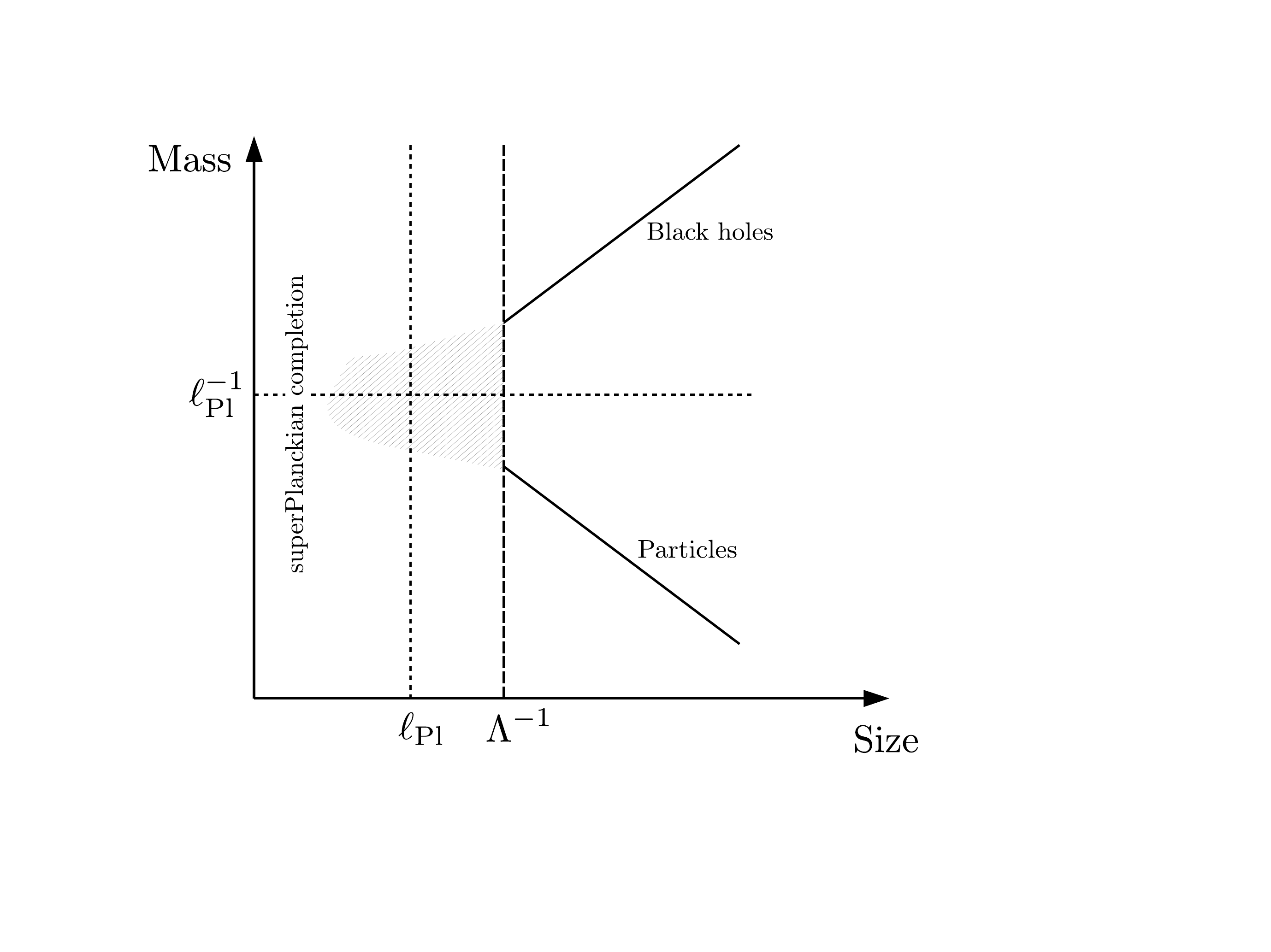}
\caption{ Sketch of physical states near the Planck scale. The graviton + matter QFT and semiclassical black hole physics both fail at short distance scales below $1/\Lambda$, which is some length greater than the Planck length ${\ell_{\rm Pl}} \equiv \sqrt{\GN}$. 
 The hatched area within the mass range  $ \sim \Lambda-\left({\ell_{\rm Pl}}^2 \Lambda\right)^{-1}$ may contain  UV states leaving an imprint in the sub-Planckian effective QFT, including global symmetry violation.
\label{fig:cartoon}
}
\end{figure}

Going higher in energy, the EFT of gravitons breaks down at energies near the Planck scale. Some new phenomena must occur at such energies to prevent unitarity violation. Perhaps black holes classicalize all amplitudes in a yet unknown aspect of nonperturbative QFT, or perhaps a completely different framework such as string theory has to supersede the general relativity Lagrangian. Provided it describes virtual black holes, such a theory would then constitute a UV completion of quantum gravity.

In summary our current understanding of sub-Planckian physics 
 of the gravity/matter system is an effective QFT of graviton and matter, together with semiclassical black hole physics. 
As illustrated in Fig.~\ref{fig:cartoon}, particles and black holes tend to become similar near the Planck scale, and new effects from the UV completion may become important. Hence  one should conservatively expect that these theories remain accurate only down to a distance scale somewhat above the Planck length. The validity scales of the particle EFT and of the classical black hole description may be somewhat different. In the following, we will often take them equal to a a single cutoff length scale $1/\Lambda$.

The ``graviton+matter" EFT should contain sub-Planckian effects originating from the Planckian completion. 
Indeed, by definition, the EFT is determined by matching a (yet unknown) QG amplitude to its EFT counterpart. Just like in any particle physics EFTs, the full quantum gravity amplitude should be reproduced on the EFT side  by diagrams involving 
the light degrees of freedom ({\it i.e.} gravitons and matter fields) interacting through general relativity vertices, together with \textit{new} diagrams involving local operators that are not predicted by Einstein gravity and which are the imprint of the UV completion.

An instance of this feature arises when considering a global symmetry, denoted $G$, which  happens to be the main focus of this work. The quantum gravity amplitude should violate global charge, at the very least through processes involving near-Planckian black holes (just like in the standard thought experiment based on a macroscopic black hole, see Sec.~\ref{se:intro}), and possibly via a more direct contact between global charges and Planckian physics. 
 On the other hand, an EFT of graviton and matter can certainly have exactly conserved global charges \textit{a priori}. Therefore a violation of $G$ on the EFT side must be described by extra local operators present in the EFT. It is worth noticing that this reasoning based on matching of the EFT to the quantum gravity completion is independent of how amplitudes are computed in the latter (\textit{e.g.} whether a QFT loop expansion applies).

 In summary, we should expect the sub-Planckian effective field theory arising from quantum gravity to take the form
\be
Z=\int {\cal D}{\rm[ fields]} e^{i S_{\rm eff}}\,~\quad 
S_{\rm eff}=  \int d^4 x \sqrt{-g} \left(
{\cal R} + {\cal L}_{G}+{\cal L}_{ \not G}+ \ldots
 \right) \label{eq:Seff}
\ee

Here ${\cal L}_{G}$ is the  pure matter effective Lagrangian respecting $G$, while the  ${\cal L}_{\not G}$ Lagrangian contains $G$-violating operators---possibly of higher dimension. 
The ellipses refer to possible modified gravity contributions such as $f({\cal R})$ and to mixed gravity-matter terms. The latter terms may possibly be $G$-breaking, however they are of higher dimension as compared to ${\cal L}_{ \not G}$ and lead thus to subleading effects in the amplitudes of the EFT.

Such arguments show how the general lore that quantum gravity violates global charge is realized at the level of the sub-Planckian effective theory, but do not \textit{quantify} the amount of global charge violation---exploring this question is the main goal of this paper.

\section{Symmetry violation from black holes and local operators}
\label{se:thermal}

Let us investigate how to quantify global symmetry violation from gravity. 

We have reviewed in \Sec{se:intro} the argument that black holes inevitably violate global symmetries. On the other hand, the general relativity Lagrangian itself does not violate global symmetries. One way to explain this feature is to realize that black holes, being thermal systems, have some access to Planckian physics which should be $G$-violating. Hence black holes can be seen as an environment where light states can come into contact with $G$-violating Planckian physics. However, even in the absence of black holes, nothing in principle forbids local operators from directly coupling light states to this $G$-violating physics.
 We may expect that in a generic setting the global symmetry would be violated faster by processes involving direct contact with Planckian physics compared to those involving the appearance of black holes.


In terms of sub-Planckian EFT defined in \eqref{eq:Seff}, this suggests that global symmetry violating effects from ${\cal L}_{\not G}$ should be at least as fast as those from putative quantum black holes. However, we cannot estimate the effect of the latter without knowledge of the full quantum gravity theory. Thus we cannot immediately draw any quantitative conclusion about ${\cal L}_{\not G}$. 

What do we gain by taking the theory at finite temperature? 
 In the thermal bath, all species, including \textit{classical} black holes, are statistically existing and interacting. This includes species heavier than the temperature scale, whose distribution is Boltzmann-suppressed but strictly nonzero. Furthermore, any allowed scattering processes $i$ will occur with some finite thermal rate per spacetime volume $\Gamma_i$, including any processes which violate the global symmetry $G$, for which we will denote the rate by $\Gamma^{\not G}_i$. If thermal equilibrium is perturbed by the introduction of some global charge asymmetry, then the $\Gamma^{\not G}_i$ will enter into the Boltzmann equations describing the dissipation of global charge (see \Sec{sec:charge}).
 
 
At finite temperature, black holes described by standard Schwarzschild solutions still exist (apart from the extreme case where the black hole
radius approaches the curvature scales induced by the thermal bath, which is irrelevant for our analysis). In fact, the thermal bath will contain classical black holes as local fluctuations. The behavior of these states is relatively well-understood compared to the zero-temperature effects of putative quantum black holes. Namely, thermal black hole production and particle capture are well described using semi-classical considerations and insights from macroscopic black hole physics, and decay is described by Hawking radiation. In addition, because of thermal equilibrium, an accurate knowledge of some of these processes is not even needed, knowing their order of magnitude is enough to ensure that the equilibrium distribution is reached. We will denote the rate of thermal processes in which  black holes induce  $G$-violation as $\Gamma^{\not G}_{\rm BH}$.

 \begin{table}
\center
\begin{tabular}{|c|c|c|}
\hline 
&     \begin{tabular}{@{}c@{}} Black holes\\ {\small $(R\geq 1/\Lambda)$} \end{tabular}
&     \begin{tabular}{@{}c@{}} UV states\\  {\small (size~$< 1/\Lambda$)} \end{tabular}
 \\ 
\hline 
    \begin{tabular}{@{}c@{}}Non-local contribution to $\Gamma^{\not G}$ \\ {\small (Boltzmann suppressed)} \end{tabular}
 & $\Gamma^{\not G}_{\rm BH}$ & 
    \begin{tabular}{@{}c@{}} $\Gamma^{\not G}_{\rm UV,\,nonloc}$  \\ {\small (\textit{e.g.} on-shell particle)} \end{tabular}  
  \\ 
\hline 
    \begin{tabular}{@{}c@{}}Local contribution to $\Gamma^{\not G}$ \\ {\small (Described by ${\cal L}_{\not G}$ in the EFT)} \end{tabular}
 &     \begin{tabular}{@{}c@{}} Unknown \\ {\small (\textit{i.e.} ``quantum black holes'')} \end{tabular}  & 
    \begin{tabular}{@{}c@{}}  $\Gamma^{\not G}_{\rm UV,\,loc}\equiv \Gamma^{\not G}_{\rm EFT} $  \\  \end{tabular}   
 \\ 
\hline 
\end{tabular} 
\caption{
Rates of global symmetry violation at sub-Planckian temperature $T<\Lambda$ expected from the quantum gravity theory. From general considerations it is expected that $\Gamma^{\not G}_{\rm BH} \ll \Gamma^{\not G}_{\rm UV,\,nonloc}\ll \Gamma^{\not G}_{\rm UV,\,loc} $.
\label{tab:terms}}
\end{table}


Let us now consider the effect of UV physics at scales smaller than $1/\Lambda$, especially in the hatched area in Fig.~\ref{fig:cartoon}. Let us assume that this near-Planckian physics has some properties similar to those of particles, such that a few basic intuitions on its behaviour still apply.
The effects from the unknown UV states  can be either local or nonlocal. A nonlocal effect occurs when an on-shell UV state is involved, and the associated rate of $G$-violation, noted $\Gamma^{\not G}_{\rm UV,\,nonloc}$, is then Boltzmann suppressed.\,\footnote{
Using the discussion from the next paragraphs,   nonlocal processes can be defined as any processes which are not described by the low-energy EFT. Details of nonlocal processes and on the nature of the Planckian states are not relevant here, the key feature required for our reasoning to apply is that these processes are Boltzmann-suppressed. }
Since the mass of these  UV states is within the range $\Lambda-1/G_{\rm N}\Lambda$, the Boltzmann suppression for such contribution is weaker than for the real black holes. Therefore we should expect  \be\Gamma^{\not G}_{\rm BH} \lesssim \Gamma^{\not G}_{\rm UV,\,nonloc}\,. \ee

The unknown UV states can  also induce local contributions to the $G$-violating rate.  These effects are precisely those encoded by local operators in the EFT in which the UV states are integrated out.  On general grounds we should expect the local contribution to be much larger than the nonlocal one since it has no Boltzmann suppression, hence\,\footnote{This is the reason why an EFT can be safely put at finite temperature $T<\Lambda$. Nonlocal contributions from on-shell heavy states are Boltzmann suppressed and their effects are thus negligible with respect to local effects. 
This is easily exemplified with a resonant thermal process $1~2\rightarrow 3~4$ \be
\int \frac{d^3p_1}{(2\pi)^3}\int \frac{d^3p_2}{(2\pi)^3} f_1 f_2 
\frac{d^3p_3}{(2\pi)^3}\int \frac{d^3p_4}{(2\pi)^3} |{\cal M}|^2 (2\pi)^4 \delta^{(4)}(p_1+p_2-p_3-p_4)\,,
\ee
with $|{\cal M}|^2=|y|^2[(s-m_{X}^2)^2-m_{X}^2 \Gamma_{X}^2]^{-1} $, where $y$ is a coupling constant and $m_{X}$ and $\Gamma_{X}$ are the mass and width of the resonance $X$.
The $f_i$ are the distribution functions, $f_i(E)=1/(e^{E/ T}\pm 1)$. 
When the temperature satisfies $T\ll m_{X}$, the main contribution to the integral is in the non-relativistic regime, for which $|{\cal M}|^2\approx |y|^2/m_{X}^2$. This can be described by a local operator in an EFT with cutoff below $ m_{X}$ . The nonlocal contribution from the resonance is Boltzmann suppressed by a factor $\sim e^{-m_{X}/T}$, much smaller that the local contribution. 
} 
\be\Gamma^{\not G}_{\rm UV,\,nonloc} \ll \Gamma^{\not G}_{\rm UV,\,loc}\,.\ee
What is $\Gamma^{\not G}_{\rm UV,\,loc}$ in the global picture  drawn in Sec.~\ref{se:general}? 
In Sec.~\ref{se:general} we have introduced the effective QFT describing gravitons and matter at distance scales above $1/\Lambda$, Eq.~\eqref{eq:Seff}. The local rate $\Gamma^{\not G}_{\rm UV,\,loc}$  simply matches  the rate $\Gamma^{\not G}_{\rm EFT}$ induced by the local operators from the ${\cal L}_{\not G}$ piece of the effective Lagrangian,
\be\Gamma^{\not G}_{\rm UV,\,loc}\equiv\Gamma^{\not G}_{\rm EFT}\,.\ee

These observations are summarized in Tab.~\ref{tab:terms}. 
In the following we will not be interested in $\Gamma^{\not G}_{\rm UV,\,nonloc} $. Our goal is rather to put a bound on $\Gamma^{\not G}_{\rm EFT}$, which will then imply bounds on the effective operators in ${\cal L}_{\not G}$.

From the above considerations, we conclude that a strong hierarchy between  $\Gamma^{\not G}_{\rm BH} $ and $ \Gamma^{\not G}_{\rm EFT}$ can be expected. The inequality is expected to hold for \textit{any} temperature  $T<\Lambda$. The interest of considering the theory at finite temperature  now becomes clear:  
 The semiclassical black hole processes $\Gamma^{\not G}_{\rm BH} $ being known, at least at the order of magnitude level, they provide a  \textit{lower bound} on  $\Gamma^{\not G}_{\rm EFT}$. The lower bound would be zero at zero temperature, but is nonzero at finite temperature, thereby providing a lower bound on global symmetry violating operators.   
 
Physically, the  $\Gamma^{\not G}_{\rm BH}\ll \Gamma^{\not G}_{\rm EFT}$ inequality embodies the intuition that direct interactions with the $G$-violating UV physics should be faster than $G$-violation via a black hole process. 
 This bound is well motivated, but is not proven either. In the following we will take it as a conjecture, which will be spelled out more precisely in Sec.~\ref{se:swamp}. Before taking this step, we  have to define a thought experiment (Sec.~\ref{se:thermo}) and we have to study the $G$-violating rates on both sides of the inequality (Secs.~\ref{se:BHs} and \ref{se:EFT}).

\section{The thought experiment: thermodynamics and stability}
\label{se:thermo}

Our basic proposal to measure global charge violation at finite temperature is the following. We put a region of spacetime in the thermal state, then inject global charges and measure how fast they disappear. Before going into the details of these processes, thermodynamics and stability of the experimental setup itself require attention. 

\subsection{A hot static universe}
\label{se:HFS}

The minimal setup of the thought experiment is to  put the whole Minkowski spacetime in the thermal state, where the energy density from relativistic species including the ones carrying  the global charge is $\rho=g_*\frac{\pi^2}{30} T^4$. 
Two difficulties immediately arise in this case. 

First, it has been long known that thermal flat (non-expanding) space is unstable due to the nucleation of large, growing black holes. In particular, Schwarzschild black holes with temperature lower than $T$ (\textit{i.e.} size larger than a critical radius $R_c \equiv (4 \pi T)^{-1}$ and mass greater than $M_c \equiv (8\pi \GN T)^{-1}$) will absorb radiation faster than they Hawking radiate and thus grow to engulf the whole space. The nucleation rate for such black holes has been found to be \cite{Gross:1982cv}
\be
\Gamma_{\rm nuc} \sim T G_{\rm N}^{-3/2} \exp\left({-\frac{1}{16\pi G_{\rm N} T^2 }}\right)\,, \label{eq:nuc}
\ee
with the restriction $T \lesssim \mPl \equiv (8\pi \GN)^{-1/2}$ to treat gravity semiclassically.  The global symmetry violating rates $\Gamma^{\not G}$ which we compute in thermal flat space are only meaningful if they are much faster than the instability driven by this nucleation rate:
\be
\Gamma^{\not G} \gg \Gamma_{\rm nuc} \label{eq:condBH} \,.
\ee
At sufficiently high temperature, the above constraint could fail for all symmetry-violating processes. In such case a black hole nucleates and absorbs the whole universe before the symmetry-violating processes we consider can occur, preventing any conclusion about the relative rates of these processes. 
This implies an upper bound on the temperature which we can consider within our thought experiment (see Sec.~\ref{se:bound}). 



Second, the energy density being nonzero, the universe expands with a Hubble rate $H\sim \sqrt{G_{\rm N}} T^2 $. However we will want to consider a system at constant temperature over an exponentially long period  of time, typically much larger than this Hubble time $1/H$. Hubble expansion would quickly decreases the temperature of the bath, hence it is an undesirable feature in the experiment.  To prevent such expansion, we can always construct a static universe solution by introducing positive FRW curvature and a cosmological constant CC. For a static universe filled with radiation of energy density $\rho = g_*\frac{\pi^2}{30} T^4$, we need a CC $\Lambda = 8 \pi \GN \rho$, resulting in a universe with curvature radius
 \be
R_\mathrm{curv} = \sqrt{\frac{3}{16\pi \GN \rho}} = \sqrt{\frac{45}{g_*}} \frac{\mPl}{\pi T^2} \,.
\label{eq:Rcurv}
 \ee
A radiation-filled static universe is stable against local perturbations~\cite{Harrison:1967zza, Barrow:2003ni}, so it is appropriate to consider the solution on long time scales. Provided that $T$ is well below the Planck scale, the finite size for the static universe is much larger than the other scales we will be concerned with, so we may treat space as Minkowski for most our analysis (see however \Sec{sec:fluc} for a case where the finiteness of the universe is relevant).

\subsection{On thermal cavities}
\label{se:cavity}

 It can be interesting to go beyond the minimal thought experiment described above, investigating ways to avoid the limit on temperature  from black hole nucleation. Allowing higher temperature would ultimately lead to a stronger bound on local operators.

A possible route is to  consider a thermal bath confined in a box. Thermodynamics of gravity  inside a spherical cavity (\textit{i.e.} inside a spatial $S_2$) has been insightfully  studied in \cite{York:1986it}. Namely, in the canonical ensemble,  two black hole solutions  in thermal equilibrium with the boundary are allowed, \textit{i.e.} have no conical defects. 
These are an unstable black hole with radius $R_c$  and a stable one with larger radius $R=r(1-(4\pi r T)^{-2})>R_c$, which approaches the cavity radius $r$ without  engulfing it. The free energy of the  stable big black hole solution competes with the one of hot flat space, providing a well defined thermodynamic statement  on whether hot flat space or the big black hole is the ground state.

It turns out that the black hole  phase is  fully excluded for $r T< 27/(32 \pi)$. However for such small cavity  the particle's momenta are highly quantized in all directions of the cavity, hence a standard thermodynamical approach does not apply. But for processes with center-of-mass energy much higher than $T$, including black hole formation by 2-body collisions (see Sec.~\ref{se:BHs}),   the distribution of   momenta remains nearly continuous at such high scale and so the effect of the cavity size is negligible. The reasoning above applies analogously to $AdS_4$, in which case the ``box" is the full spacetime and the black hole phase transition is the usual Hawking-Page transition.

Further developments on thermodynamics and small black holes in cavity   would be certainly worthwhile. 

\section{Thermal black holes}
\label{se:BHs}

Having discussed the limits to the stability of hot space, we are now ready to consider the population of black holes in a thermal bath, which will act to violate global charge. 

Let us first observe that in exact thermal equilibrium black holes of mass $\MBH$ (which have entropy $S = 4 \pi \GN \MBH^2$) would appear in the thermal bath with density $\propto \exp(-\MBH/T + 4 \pi \GN \MBH^2 )$. The Boltzmann factor $\exp(-\MBH/T)$ suppresses the population of heavier black holes, but for $\MBH \gtrsim 1/(4\pi \GN T)$ (\textit{i.e.} $\RBH \gtrsim 1/2\pi T$ ) the entropy factor in fact dominates over the Boltzmann suppression and the thermal density diverges with increasing $\MBH$, which is clearly unphysical. And indeed, as discussed in Sec.~\ref{se:HFS} we know that black holes of radius greater than $R_c = 1/4\pi T$ grow in the thermal bath instead of decaying, so their presence triggers the phase transition away from hot flat space. 

Thus we actually want to consider a setting in which small black holes exist in a thermal bath, but black holes with $\RBH > R_c$ do not. This is sensible if we imagine starting from a thermal bath of ordinary QFT particles (\textit{i.e.} no black holes initially), and are able to show that a population of small black holes is thermally produced before the system is swallowed up by unstable large black holes. To verify this we will consider various processes which produce black holes in a thermal bath and estimate their rates (on a logarithmic scale). 

Given that the unstable large black holes do not have time to form, the most common black holes in the thermal distribution will be those with the smallest possible mass and radius. This will be determined by a cutoff scale $\Lgrav$ at which the validity of semiclassical Einstein gravity breaks down; we can only reliably consider black holes with radius greater than $R_* \sim \Lgrav^{-1}$ and therefore mass greater than $M_* = R_*/2\GN \sim 4\pi \mPl^2/\Lgrav$. This $\Lgrav$ could for example be the string scale or the size of some compact extra dimensions. \textit{A priori} $\Lgrav$ could be as large as the Planck scale where quantum gravity effects are necessarily strong, though we will generally assume at least a small hierarchy between $\Lgrav$ and $\mPl$ to ensure control. 

\subsection{Boltzmann equation for black hole formation}

Based on semiclassical arguments one expects that two colliding  particles with super-Planckian center-of-mass energy $\sqrt{s}$ should form black holes with a cross section corresponding to the classical black hole area $R \sim G_{\rm N} \sqrt{s}$~\cite{Banks:1999gd, Giddings:2001bu, Dimopoulos:2001hw}. From this cross-section we can obtain a thermal formation rate for black holes from two-body collisions. This rate has been previously discussed in Refs.~\cite{Conley:2006jg, Borunda:2009wd, Nakama:2018lwy}.

The 2-body collision term in the Boltzmann equation for number density is in general given by
\be
\int\frac{d^3p_1}{(2\pi)^3} \int\frac{d^3p_2}{(2\pi)^3} f_1 f_2\, \langle\sigma(s,t,u) v \rangle \equiv
\int\frac{d^3p_1}{(2\pi)^32E_1} \int\frac{d^3p_2}{(2\pi)^32E_2} f_1 f_2 \gamma(s,t,u)
\ee
where on the right-hand side we make appear the Lorentz-invariant phase space. $\gamma(s,t,u)$ is a dimensionless rate, usually given in QFT by the integral  of $|{\cal M}|^2$ over the final states phase space. For black hole production, assuming a minimum possible black hole mass of $M_*$ as defined above, we have
\be
\gamma_2(s) \approx  \frac{\pi}{4} G_{\rm N}^2 s^2\, \Theta(s>M^2_*)\,.
\ee 
This quantity is, in a sense, simpler and more fundamental than a scattering cross-section. The $f_1$, $f_2$ statistics give Boltzmann factors in the integrand, such that the rate is dominated by minimal total energy that can produce a black hole, $E_1 \approx E_2 \approx M_*/2$. Since these energies are super-Planckian, the particles can always be treated as ultrarelativistic, and the total rate of BH formation is given by
\be 
 \Gamma^{\rm rel}_2\approx \frac{\kappa}{128\pi^3} G_{\rm N}^2 \int dE_1 E_1\int dE_2 E_2 \int d\cos\theta \, e^{-(E1+E2)/T} \,s^2\, \Theta( s > M_*^2 )\,
 \ee
with $s=2E_1 E_2(1-\cos \theta)$. For $T\ll M_*$, all the integrals can be done and give 
\be
\Gamma^{\rm rel}_2\approx \frac{\kappa}{128\sqrt{2}\pi^{5/2}} G_{\rm N}^2 \sqrt{M_*^{11} T^5} e^{-M_*/T}\,.
\ee
The above rate is independent of the species of particles which are colliding, so every pairwise combination of species would contribute to the total rate.

\subsection{Black holes from thermal fluctuations}
\label{sec:fluc}

So far we have described black hole formation in terms of scattering processes, just as we would compute the formation rate of a heavy particle. However in a thermal system we may take a more general approach of considering black holes as being formed by large thermal fluctuations of local energy density. This effectively counts any different ways in which a black hole could form and is thus enhanced by an exponentially large entropy factor. We must check that these entropy factors do not favor the production of large black holes that destabilize the thermal bath (those with $\RBH > R_c \sim 1/T$) 

Piran and Wald~\cite{Piran:1982pp} computed a rate for black hole formation from fluctuations of local energy density in a box of thermalized radiation. They assumed that a black hole of size $R$ forms if the energy of radiation in some local volume of size $\sim R^3$ exceeds $\sim R/\GN$. Their result for the rate to form a black hole of mass $M$ and size $R$ can be written as 
\be
\Gamma^\mathrm{rel}_\mathrm{fluc} \sim \frac{1}{R^4} \exp\left(-M/T + S_\mathrm{fluc} \right)
\label{eq:PWRate}
\ee
where
\be 
S_\mathrm{fluc} \sim (M R)^{3/4} \sim \GN^{3/4} M^{3/2} \,.
\ee 

The first term in the exponential is a Boltzmann suppression for realizing a fluctuation of energy $M$, while $S_\mathrm{fluc}$ represents the entropy associated with this fluctuation (a volume of relativistic gas of total energy $M$ and volume $R^3$). Importantly, this entropy is \emph{not} the black hole entropy $S = 4\pi \GN M^2$, but instead is always smaller (which is to be expected from entropy bounds). Furthermore, in the regime $R < 1/T$, for which the resulting black hole decays instead of growing, the entropy $S_\mathrm{fluc}$ is subdominant compared to the Boltzmann factor $M/T$. 

The entropy factor $S_\mathrm{fluc}$ would appear to dominate for $M \gtrsim \GN^{-3/2} T^{-2}$, which corresponds to $R \gtrsim \GN^{-1/2} T^{-2}$. However, at these scales the curvature of space becomes relevant. As discussed in \Sec{se:HFS}, for our thought experiment we consider static universes, which are closed with radius of curvature $\sim (8 \pi \GN)^{-1/2} T^{-2}$. Thus the result in \Eq{eq:PWRate} necessarily breaks down before we reach volumes of this size; the black hole with $S_\mathrm{fluc} > M/T$ cannot ``fit" into the finite static universe. Therefore, within our setup the black hole production rate is always dominated by the Boltzmann factor $\exp(-M/T)$ for relevant masses $M$.

The analysis of \Ref{Piran:1982pp} was restricted to the case of a thermal gas of massless particles. One may ask whether the conclusions are altered in the case of non-relativistic particles. Note that even if we consider a gas of particles with mass $m > T$, within the fluctuation forming a black hole the typical temperature is $(M/R^3)^{1/4} \sim R^{-1} \sqrt{\mPl R}$. Thus the fluctuation only behaves non-relativistically if we have $m > R^{-1} \sqrt{\mPl R} \gg R^{-1}$. In this regime the fluctuation appears as a gas of point-like massive particles, and we can make a simple probabilistic estimate of its rate of appearance. The average number density of particles of mass $m$ is $\langle n \rangle = \left( \frac{m T}{2 \pi} \right)^{3/2} \exp(-m/T)$. Treating the particles as independent (classical and non-interacting), the probability to find a certain number of particles in a volume $\sim R^3$ will obey a Poisson distribution with mean $\langle n \rangle R^3$. A black hole of total mass $M$ will certainly form when a total number of particles $N = M/m$ occupy a spherical volume of radius $R \sim 2 \GN M$. We can express this probability as  
\be
{\rm P}_{\rm BH} =
e^{-\langle n \rangle R^3} \frac{\left(\langle n \rangle R^3\right)^{N}}{N!}\,. \label{eq:PBH}
\ee
A  formation rate per unit volume and time would be obtained by dividing by $V$ and by a timescale for classical black hole formation. As in the relativistic case we are most interested in the behavior on a logarithmic scale; taking the log gives 
\be
\log \Gamma^\mathrm{non-rel}_\mathrm{fluc} \sim -M/T + N \log \left[ \frac{R^3}{N} \left( \frac{mT}{2\pi} \right)^{3/2} \right] + N - \langle n \rangle R^3 \,.
\ee
In the first term we recognize the Boltzmann factor, and in the second term the entropy of a non-relativistic gas of $N$ particles in volume $R^3$ as given by the Sackur-Tetrode equation. This result thus closely parallels that for the relativistic case, \Eq{eq:PWRate}. In terms of the mass $M$ of the black hole we have 
\be
\log \Gamma^\mathrm{non-rel}_\mathrm{fluc} \sim -\frac{M}{T} + \frac{M}{m} \left( \log \left[ \GN^3 M^2 m \left( \frac{mT}{2\pi} \right)^{3/2} \right] + 1 \right) - \langle n \rangle \GN^3 M^3
\ee
Since we are working in the regime $m \gg T$, we see that the second term corresponding to entropy enhancement is sub-dominant to the Boltzmann factor except when $M$ becomes exponentially large ($\propto \exp \frac{m}{2T}$). But as before, in our thought experiment the black hole size is restricted by the finite size of the static universe $\sim (8\pi \GN^{-1/2}) T^{-2}$ so these large masses cannot be realized.

We conclude that within our thought experiment black hole formation is never enhanced exponentially by factors which can dominate over the Boltzmann suppression $e^{-M/T}$. Thus the smallest black holes, of mass $M_* \sim (\GN \Lambda)^{-1}$, are always produced exponentially faster than heavier ones which can destabilize the space.

\subsection{Equilibrium distributions and thermalization time}

Having estimated the rate of black hole formation, let us return to considering the equilibrium distribution of black holes and the the time required to realize them.

We can treat black holes of particular mass $M$ as classical heavy particles with large degeneracy due to the BH entropy, which counts microstates. So in equilibrium the number density of black holes with mass between $M$ and $M+ dm$ is
\be
\frac{dn_{\rm BH}}{dM} dM \sim g_{\rm BH}(M)\, e^{-\frac{M}{T}} \,\left(\frac{M T}{2\pi}\right)^{3/2} dM  \,. \label{eq:nBH0}
\ee
Here $g_{\rm BH}$ represents a density of states for black holes of mass $M$, which is related to their thermodynamic entropy by $g_{\rm BH} \sim e^S/\mPl$~\cite{York:1986it, Braden:1987ad}. So the thermal number density goes as 
\be
\frac{dn_{\rm BH}}{dM} dM \sim \GN \sqrt{M^5 T^3} \exp(-M/T + 4\pi \GN M^2) dM  \,. \label{eq:nBH}
\ee

The thermalization time required to achieve this number density is obtained by dividing by the BH formation rate $\Gamma(M)$. In the above we found that $\log \Gamma(M) \sim -M/T + S_\mathrm{fluc} + \cdots$ where $S_\mathrm{fluc}$ was much less than the black hole entropy in relevant cases. Thus the thermalization time to achieve the density \Eq{eq:nBH} is simply
\be
\tau_\mathrm{therm}(M) \propto n_\mathrm{BH}(M)/\Gamma(M) \sim \exp(4\pi \GN M^2)
\label{eq:tautherm}
\ee
Importantly, because the black hole formation rate is not enhanced by the black hole entropy, large black holes do not thermalize faster.

As discussed, the equilibrium density decreases as $M$ is increased from the minimal value $M_*$, until we go above $M_c \sim 1/ 4\pi \GN T $ where it starts to diverge signalling the instability of the system. However, the formation rate of these black holes, $\propto \exp(-1/(4\pi \GN T^2)$, is always less than the thermalization time \Eq{eq:tautherm} for black holes much smaller than $M_c$. The  same is true of the nucleation rate \Eq{eq:nuc} computed in~\Ref{Gross:1982cv}. 

We are therefore justified in only considering small black holes with $M \ll M_c$ in the thermal density \Eq{eq:nBH}. This density is dominated by the smallest black holes of mass $M_*$, so after integrating over mass we obtain a total number density for black holes of
\be
n_{\rm BH} \sim \GN (M_*T)^{5/2}
 e^{-M_*/T+4\pi \GN M_*^2}\,. \label{eq:nBHeq}
\ee
Again, the entropy factor is always subleading since $M_* \ll M_c \sim (\GN T)^{-1}$.

Although the entropy factor is subleading, it will be important in the context of global symmetry violation, as we will see in Sec.~\ref{se:bound}. 

\subsection{Particle capture}
\label{se:capture}

Once a population of microscopic black holes is established, they will absorb particles in the thermal bath. The most common black holes are those with the minimal size $R_*$, which is much smaller than the typical wavelength of particles in the thermal bath, so that the absorption is a quantum process. The absorption cross-section for a scalar particle of mass $m$ impinging on a Schwarzschild black hole of radius $R_*$ with velocity $v$ was obtained by Unruh~\cite{Unruh:1976fm}, with approximate behavior:

\eq{
\sigma_\mathrm{capture} \approx  
\left\{ \begin{array}{ll}
\frac{8 \pi R_*^2}{v} \frac{\pi R_* m}{v}  &,\quad v \lesssim \pi R_* m   \\
\frac{8 \pi R_*^2}{v} &,\quad v \gtrsim \pi R_* m  \,.
\end{array}
\right. 
\label{eq:sigmacapture}
}

(The absorption cross-section for fermions is related to that of scalars by an overall numerical factor~\cite{Unruh:1976fm}.) For a non-relativistic particle species we have $\langle v \rangle \sim \sqrt{T/m}$, so that the upper case is relevant only for $m^3 > T/R_*^2$. Importantly, absorption of particles by black holes is a well-understood \textit{low-energy} process, unlike black hole formation in high-energy collisions for which only inclusive rates can be discussed. Absorption by black holes is therefore a reliable indicator of the destruction of global charge.

The absorption rate for a given species $X$ by the micro black holes appearing in the number density Boltzmann equation of $X$ is given by
\be
\frac{dn_X}{dt}=- n_X n_{\rm BH} \langle \sigma v  \rangle \,,
\ee
where $n_{\rm BH}$ is set by  the equilibrium value Eq.~\eqref{eq:nBH} and $\langle \sigma v  \rangle$ indicates a thermal average. This has a particularly simple form when the second case of \Eq{eq:sigmacapture} applies: 
\be
\Gamma_{\rm capture} \approx 8 \pi n_X n_{\rm BH}\, R_*^2.
\label{eq:Rcap}
\ee
For the case where $\langle v \rangle \lesssim \pi R_* m$, there is an additional factor, though of course in either case the most important factor suppressing the rate is the exponentially small BH density $n_{\rm BH}$.

\subsection{Discussion}

Let us summarize what we have learned from the above considerations of thermal black holes. We have argued that we can sensibly consider a ``superheated'' static universe on time scales for which small black holes (of radius corresponding to a cutoff $R_* \sim \Lambda^{-1}$) have achieved a thermal density $\propto \exp(-M_*/T)$, while black holes large enough to destabilize the universe (those with radius greater than $R_c = (4\pi T)^{-1}$) have not had time to be produced. This separation of course requires that $R_* < R_c$, meaning $T \lesssim \Lambda/(4\pi)$, \textit{i.e.} the temperature must be well below the Einstein gravity cutoff.    

Processes involving black holes are expected to violate global symmetries, though such effects may not be reflected in semiclassical descriptions. For example, super-Planckian scattering of particles may violate global charge conservation in the process of forming a black hole, though we cannot describe this without a detailed quantum gravity model. Out of all the processes considered in this section, the highest rate corresponds to capture of particles by black holes (\Eq{eq:Rcap}), since this is enhanced by the black hole entropy. This is convenient since the capture is a low-energy process which is not affected by details of quantum gravity. Capture of globally charged particles is thus a well-controlled violation of global symmetry, and we can write the rate of global charge violation from black hole effects as
\be
\Gamma^{\not G}_{\rm BH}=\Gamma_{\rm capture}+ \ldots \label{eq:BH_rate}
\ee
where the ellipses correspond to the sub-dominant effects from black hole formation. Throughout this work we will be interested in logarithmically comparing this rate to that of other global symmetry violating processes. The dominant term in the log rate is simply the Boltzmann factor: 
\be
\log \Gamma^{\not G}_{\rm BH}(T) \approx -\frac{M_*}{T} = -\frac{1}{2\GN \Lambda_\mathrm{grav} T}\,. \label{eq:log_BH_rate}
\ee

\section{Global symmetry violation from effective operators }

\label{se:EFT}

In this section we consider global symmetry violation by local operators on the EFT side, our goal being to evaluate the  $\Gamma_{\rm EFT}^{\not G}$ rate.

For our purposes it is enough to focus on scalar particles---the approach applies similarly to fields with higher spin. We consider a complex scalar $\phi_\pm$ charged under $U(1)_G$. 
The effective Lagrangian contains in principle an infinity of operators built from monomials of $\phi_\pm$  of increasingly  high dimension. By definition the contribution of these operators to physical processes decrease with the operator dimension. For our purposes we consider a single operator with a generic parametrization, which can be understood as being among the dominant ones in a given EFT. 

The $G$-violating local operator we consider  take schematically the form
\be
 {\cal O}_{\not G}=\frac{\partial^{2r} (\phi_+)^{p}(\phi_-)^{n_{\not G}-p} }{p!(n_{\not G}-p)!}
 \,,\quad {\rm with}  \quad n_{\not G} \neq 2p\,, 
\ee
and enter in the effective Lagrangian as
\be
 {\cal L}_{\rm EFT} \supset c_{\not G} \frac{\partial^{2r} (\phi_+)^{p}(\phi_-)^{n_{\not G}-p}}{p!(n_{\not G}-p)!\,\Lambda^{2r}\,\mu^{n_{\not G}-4}} +{\rm h.c.} 
 \label{eq:OGv}
\ee
The dimension of the operator is  $d_{\not G}=2r+n_{\not G}$, and the operator violates $G$ by $\Delta G=n_{\not G}-2p $. $\Delta G$ is defined up to a sign, hence it must appear as $|\Delta G|$ in physical quantities. 

Regarding the scales, our convention is that $\Lambda$ is the EFT cutoff scale and  $ 4\pi \mu=\Lambda$. 
These scales are  set by the presence of other, $G$-respecting higher-dimensional operators assumed to be present in the effective Lagrangian, following the same-power counting as Eq.~\eqref{eq:OGv} but with a $c=O(1)$ coefficient.  The coefficient $ c_{\not G}$ is a free parameter we want to estimate, and may possibly be much smaller than one.

 Many inequivalent operators potentially exist because of the many combinations of derivatives---even though some combinations cam be removed using equations of motion at a given order \textit{i.e.} field redefinitions. For our purposes it is enough to focus on a single one at a time. Leaving apart the combinations of derivatives, the effective operator described in Eq.~\eqref{eq:OGv} is characterized by a few numbers which are the overall coefficient  $ c_{\not G}$, the dimension $ d_{\not G}$, the number of fields $ d_{\not G}$ and the amount of charge violation $\Delta G$.  The three last numbers satisfy
 \be
0< |\Delta G| \leq n_{\not G} \leq d_{\not G} \,.
 \ee

The general expression for the rate of change in number density appearing in the Boltzmann equation of QFT is given in Eq.~\eqref{eq:rate_gen}.
In the relativistic case, the rate of any process induced by ${\cal O}_{\not G}$ with $n$ initial states and $k=n_{\not G}-n$ final states, noted $\{n\}\rightarrow\{k\}$, is  estimated to be of order
\be
\Gamma^{\not G}_{ \rm EFT} \sim \kappa_i \kappa_f c_{\not G} \frac{T^{2d_{\not G}+4}}{\Lambda^{2d_{\not G}}} \label{eq:rate_EFT_rel}
\ee
where one has used that each phase space integral  brings a factor of $1/4\pi$. 
The $\kappa_i$,  $\kappa_f$ correspond to the symmetry factors for incoming and outgoing particles. 

In the non-relativistic case, for $T\gg m$, it turns out that the highest rates are when $n\approx k \approx n_{\not G}/2$. More initial states would imply more Boltzmann factors, while more final states would require the initial states to have enough energy to pass the kinematic threshold for production, implying again a Boltzmann suppression. 
The two effects are actually equivalent because, in the classical limit where $f_i=e^{-E_i/T}$, energy conservation implies that $\prodsym_i f_i=\prodsym_f f_f$, such that $\Gamma(\{n\}\rightarrow \{k\})=\Gamma(\{k\}\rightarrow \{n\}) $. (For relativistic particles this relation is only approximately true due to spin-dependent $O(0.1)$ corrections coming from the quantum nature of the particles.)

For non-relativistic initial and final states,  it turns out that the rate can be computed exactly.
We focus on an operator with no derivatives, \textit{i.e.} $r=0$ in Eq.~\eqref{eq:OGv}. Derivatives can be trivially  taken into account in the non-relativistic limit since $\partial_\mu \phi=m \delta^0_{\mu} \phi$. 
The calculation in the relevant case $n\approx k \approx n_{\not G}/2$ is described in App.~\ref{app:NR_EFT}.   We obtain
\be
\Gamma^{\not G}_{\rm EFT}= \kappa_i \kappa_f
e^{- n_{\not G} m  /2T} \frac{\Gamma(3n_{\not G}/2-4)}{n_{\not G}^{3/2}\Gamma(3n_{\not G}/4-2)\Gamma(3n_{\not G}/4-3/2))}
\frac{  16\pi^2 }{ (2 \pi^3)^{n_{\not G}/2}} \frac{(m T^3)^{n_{\not G}/2}}{ \sqrt{m^3 T^5} \Lambda^{2n_{\not G}-8}}  \,.
\label{eq:rate_EFT_NR}
\ee

Let us finally comment on the symmetry factors. At fixed $n_{\cal G}$, if $\Delta G \ll n_{\cal G}$, the largest combinatoric value is when  $\phi_\pm$ fields are in equal amount both in initial final state. 
The smallest value is when most of fields are either in initial or final state. This gives 
\be
\frac{1}{ (n_{\not G}/2)!^2 }<\kappa_i \kappa_f<\frac{1}{ (n_{\not G}/4)!^4 }\,.
\ee
Note that in the non-relativistic case we have considered the configuration corresponding to the upper bound. 
On the other extreme, if $\Delta G = n_{\not G}$, the values of  combinatoric factors are 
\be
\frac{1}{ (n_{\not G})! }<\kappa_i \kappa_f<\frac{1}{ (n_{\not G}/2)!^2 }\,.
\ee
Note however that logarithmically all these values are of  same order of magnitude, $\sim \nng \log \nng$. In the large $n_{\not G}$ limit we may write the following bounds on the logarithms of the rates \Eqs{eq:rate_EFT_rel}{eq:rate_EFT_NR}:
\be
\log \frac{\Gamma^{\not G}_{ \rm EFT,rel}}{\Lambda^4} \gtrsim \log \cng - \nng \big(\log n_{\not G}+O(1)\big)  + 2\nng \log \frac{T}{\Lambda} \label{eq:log_rate_EFT_rel} 
\ee
\be
\log \frac{\Gamma^{\not G}_{ \rm EFT,non-rel}}{\Lambda^4} \gtrsim -\frac{\nng m}{2T} + \log \cng - \nng \big(\log n_{\not G}+O(1)\big) + \frac{\nng}{2} \log \frac{m T^3}{\Lambda^4} \label{eq:log_rate_EFT_rel} \,
\ee
where ``$O(1)$" indicates numerical factors that do not depend on the
 parameters $\cng$, $\nng$, $T$, $\Lambda$. 

\section{Global symmetry violation in the thermal bath}
\label{se:bound}

We are finally in a position of carrying out the thought experiment. Namely, we prepare hot flat space in the thermal state, which contains a Boltzmann-suppressed distribution of black holes dominated by the smaller ones of radius $R_*$. We then inject global charges and measure how fast they disappear.

The global charges are destroyed either by thermal black holes or by local operators in the EFT. Following 
the arguments of Secs.~\ref{se:general},\,\ref{se:thermal}, one may conjecture that the  rate of $G$-violation from the local operators, $\Gamma_{\rm EFT}^{\not G}$, is at least as large as the $G$-violation rate from black holes, $\Gamma_{\rm BH}^{\not G}$. In the following we will compare these two rates and determine the conditions for which we indeed have $\Gamma_{\rm EFT}^{\not G} \gtrsim \Gamma_{\rm BH}^{\not G}$.

For our purposes it is enough to focus on $U(1)$ global symmetries. Our reasonings would necessarily apply to larger symmetries such as $U(N)$ or $O(N)$, which contain $U(1)$ as subgroups. We consider particles and antiparticles charged under a $U(1)$ group $G$ with charges $q$ and $-q$ respectively.  

\subsection{ Global charge time evolution }
\label{sec:charge}

Consider an interacting set of $n_+$ particles and $n_-$ antiparticles. The interactions respecting $G$ lead to chemical reaction rates of the form $\Gamma_{\rm EFT}^G\propto(n_+n_-)^m $ with $m\leq 0$. This immediately implies that the Boltzmann equations for $n_+$ and $n_-$ are equal,
\be
\frac{d n_+}{dt}=\frac{d n_-}{dt}= \sum_i \Gamma_{{\rm EFT},i}^G\,. \label{eq:Boltz1}
\ee

The total charge $Q$ of the system is defined by  
\be
Q=q (n_+-n_-)\,, \label{eq:Q}
\ee
which is CPT invariant. 
In the presence of $G$-respecting interactions, it follows from Eq.~\eqref{eq:Boltz1} that the Boltzmann equation for $Q$ is
\be
\frac{d Q}{dt}=0\,,
\ee
which is simply the fact that the $G$ charge is conserved along the thermal evolution of the system.

For our purposes it is sufficient---and convenient---to study an  injection of global charge which is small with respect to the equilibrium quantity, such that the total densities are given by
\be
n_{\pm}=\delta n_{\pm} +n^{\rm eq}\,,\quad \delta n_{\pm} \ll n^{\rm eq}
\ee
where $n^{\rm eq}_+=n^{\rm eq}_-\equiv n^{\rm eq}$.

Let us first consider the black hole side. Once the presence of the thermal black holes is taken into account, global charges can be captured as described in Sec.~\ref{se:capture}. 
The Boltzmann equations now contain the contribution
\be
\frac{d n_+}{dt} \supset -\left(1+\frac{\delta n_+}{n^{\rm eq}}\right) \Gamma^{\not G}_{\rm BH}\,,\quad\quad \frac{d n_-}{dt} \supset - \left(1+\frac{\delta n_-}{n^{\rm eq}}\right) \Gamma^{\not G}_{\rm BH}\,.
\ee
 The Boltzmann equation for $Q$ is now given by 
\be
\frac{d Q}{dt}= -q\frac{\delta n_+-\delta n_-}{n^{\rm eq}} \Gamma^{\not G}_{\rm BH}=
-\frac{Q}{n^{\rm eq}} \Gamma^{\not G}_{\rm BH} \,.
\ee
Hence the $G$-charge is not conserved anymore and  vanishes at the rate
\be
Q(t)= Q(t=0)\exp\left({-\frac{\Gamma^{\not G}_{\rm BH}}{n^{\rm eq}} t}\right)\,.
\ee

A similar calculation can be done for the EFT side. A single local operator ${\cal O}_{\not G}$ (see Eq.~\eqref{eq:OGv}) gives rise to many different processes, which all have the same amount of $G$ violation, given by the $\Delta G$ associated to the ${\cal O}^{\not G}$ operator. 
A given process $a \phi_+ + b \phi_- \rightarrow c \phi_+ + d \phi_-$ induced by ${\cal O}_{\not G}$ satisfies $a+b+c+d=n^{\not G}$ and $\Delta G=(a+d)-(b+c)$. 
Such process is automatically accompanied by its complex conjugate induced by the ${\cal O}^\dagger_{\not G}$. Moreover, in Boltzmann equation, these processes come together with the reverse ones under PT.  The four processes to take into account are thus 
\begin{align}
& a \phi_+ + b \phi_- \rightarrow c \phi_+ + d \phi_- \nonumber \\
& a \phi_- + b \phi_+ \rightarrow c \phi_- + d \phi_+  \nonumber \\
& c \phi_+ + d \phi_- \rightarrow   a \phi_+ + b \phi_-  \nonumber \\
& c \phi_- + d \phi_+ \rightarrow   a \phi_- + b \phi_+   ~.
\end{align}

The contributions to Boltzmann equation are then given by 
\be
\frac{d n_+}{dt} \supset 
\bigg[(a-c)\left(-a\delta n_+-b\delta n_- + c\delta n_++d\delta n_- \right)
+(b-d)\left(-a\delta n_--b\delta n_+ + c\delta n_-+d\delta n_+ 
\right)\bigg] \frac{\Gamma^{\not G}_{\rm EFT}}{n^{\rm eq}} \nonumber
\ee
 \be
\frac{d n_-}{dt} \supset 
\bigg[(a-c)\left(-a\delta n_--b\delta n_+ + c\delta n_-+d\delta n_+ \right)
+(b-d)\left(-a\delta n_+-b\delta n_- + c\delta n_++d\delta n_- \right)
\bigg] \frac{\Gamma^{\not G}_{\rm EFT}}{n^{\rm eq}} \,.
\ee

In the evolution equation of $Q$, it turns out that the rates combine such that the CPT-invariant combination $(\Delta G)^2$ appears,
\be
\frac{dQ}{dt}=-\,(\Delta G)^2\,Q \frac{\Gamma^{\not G}_{\rm EFT}}{n^{\rm eq}}\,. 
\ee
Therefore the time evolution of the total charge follows
\be
Q(t)= Q(t=0)\exp\left({-(\Delta G)^2 \frac{\Gamma^{\not G}_{\rm EFT}}{n^{\rm eq}} t}\right)\,.
\ee

\subsection{ Scales }
\label{eq:scales}

Let us sum up the information about the various scales gathered in the previous sections. Smallest black holes have radius $R_*=2G_{\rm N} M_*$.  To ensure that the  black holes have a semiclassical description without sizeable effects from the UV completion, one should require $M_*$ to be sizeable with respect to $1/\sqrt{G_{\rm N}}$. 
 To avoid nucleation of the whole universe one requires the maximum temperature of the bath $T_*$ to be 
 \be 
 T_*=\frac{1}{8\pi R_*}=\frac{1}{16 \pi G_{\rm N} M_*} \,. \label{eq:Tstar} \ee 
For this value, $\Gamma_{\rm BH}$  is still exponentially enhanced with respect to $\Gamma_{\rm nuc}$ because of the enhancement from entropy in the capture rate. We recall that this requirement  on temperature might be avoidable by devising a thought experiment preventing nucleation of big black holes  (see Sec.~\ref{se:thermo}).

  On  EFT side the temperature should not exceed the cutoff $\Lambda$, which itself should not exceed the (non-reduced) Planck mass $1/\sqrt{G_{\rm N}}$. In the massive case, the particle mass should satisfy $m<\Lambda$. 
Finally, although it is not fully mandatory, it is natural to take a single universal cutoff for both the EFT and Einstein gravity, \textit{i.e.} taking $\Lgrav = \Lambda$ and therefore setting the minimal black hole radius to $R_*=1/\Lambda$.
 
 The relation between scales can then be summarized as
\be
T\leq T_* =\frac{\Lambda}{8\pi} <\Lambda < \frac{1}{\sqrt{G_{\rm N}}}<M_*=\frac{1}{2 G_{\rm N} \Lambda}\,.
\ee
The only freedom is on the value of $\Lambda $, \textit{i.e.} the maximal scale at which we trust EFT and a semiclassical description of  black holes. It follows that the maximal rate $\Gamma^{\not G}_{\rm BH}$ for symmetry violation from thermal black holes goes as
\be
\Gamma^{\not G}_{\rm BH, max} \propto \exp\left(-\frac{M_*}{T_*}\right) \approx \exp\left(-\frac{4\pi}{\GN \Lambda^2}\right) \,.
\ee

\subsection{ Bounding ${\cal O}_{\not G}$ }

We can  finally determine the conditions for the local operator ${\cal O}_{\not G}$ to mediate global symmetry violation faster than black holes.

The  $(\Delta G)^2\Gamma^{\not G}_{\rm EFT} \gtrsim \Gamma^{\not G}_{\rm BH}$ criterion should be understood at the order-of-magnitude level. 
For example, on the EFT side, for an operator with a given dimension and number of fields, we may wonder which $\Delta G$ and which process should be picked exactly. However these are fine details which cannot be controlled by our criterion, and which are irrelevant because they constitute  only small corrections to the result obtained. 
The bounds are conveniently treated logarithmically. In the formulas shown below, we drop  small log corrections and  assume  $n_{\not G}$ is large so that we  show only the resulting leading term.

 In the relativistic case, the bound is set by comparing Eqs.~\eqref{eq:BH_rate} and \eqref{eq:rate_EFT_rel}.  The most stringent condition is achieved for $T=T_*$, such that 
\be
 \log c_{\not G}  - n_{\not G} \big(\log n_{\not G}+O(1)\big) - d_{\not G} \log\left(\frac{\Lambda}{T}\right)   \gtrsim -\frac{M_*}{T_*}=- \frac{4\pi}{G_{\rm N} \Lambda^2} \,. \label{eq:bound_rel_gen}
\ee
The last equality is set using the relations in Sec.~\ref{eq:scales}. 
For an operator of moderate dimension, the $d_{\not G}$  and $n_{\not G}$ terms in Eq.~\eqref{eq:bound_rel_gen} can be neglected and we conclude that  the operator's overall coefficient should satisfy 
\be
 \log c_{\not G} \gtrsim - \frac{4\pi}{ G_{\rm N} \Lambda^2}\,. \label{eq:bound_c}
\ee
Conversely, for $c_{\not G}=O(1)$, an upper bound on the number of fields and derivatives is obtained.
For instance, for $d_{\not G}=n_{\not G}$ case, and for large $n_{\not G}$, we obtain
\be
n_{\not G} \left(\log n_{\not G} + O(1) \right) \lesssim \frac{4\pi}{G_{\rm N} \Lambda^2}\, \label{eq:bound_n}
\ee
where one has used $T_*=\Lambda/8\pi$. In the large $\nng$ limit a sufficient condition enforcing Eq.~\eqref{eq:bound_n} is $\nng \lesssim \frac{4\pi}{G_{\rm N} \Lambda^2}/\log \frac{4\pi}{G_{\rm N} \Lambda^2}$. This bound is of interest since  gauge symmetries could potentially constrain the lowest dimension of a global symmetry violating operator to be very high (see \Sec{se:WGC}).

In the non-relativistic case the condition is obtained by comparing Eqs.~\eqref{eq:BH_rate} and \eqref{eq:rate_EFT_NR}
\be
 \log c_{\not G}  - n_{\not G} \big(\log n_{\not G}-\frac{1}{2}\log\left(
\frac{m T^3}{\Lambda^4}  
 \right)+O(1)\big) - \frac{n_{\not G} m}{2T}     \gtrsim -\frac{M_*}{T} \,.
\ee
Here we have $m \gg T$ hence $T$ is kept as a free parameter. For moderate $n_{\not G}$ and $T$ close to $m$, the  conditions on $c_{\not G}$ and $n_{\not G} $ are the same as in the relativistic case. For  $T\ll m$, we obtain a temperature-independent condition  on the mass of the fields appearing in  the ${\cal O}_{\not G}$ operator, 
 \be
n_{\not G} m \lesssim \frac{1}{G_{\rm N}\Lambda}\,. \label{eq:bound_m}
\ee

\section{A swampland symmetry conjecture}
\label{se:swamp}

In Secs. \ref{se:thermo} through \ref{se:bound} we have established a scenario in which one can quantitatively compare global-symmetry-violating effects from gravitational processes (\textit{i.e.} black holes) with those from EFT processes. Furthermore, the arguments of Sec.~\ref{se:thermal} suggest that in the sub-Planckian effective QFT, symmetry violation by black-hole-mediated processes should be slower than symmetry violation by local operators---which are understood to encode the low-energy effects from the Planckian completion. One may view such a bound as a glimpse of a general principle quantitatively constraining the global symmetries of effective field theories coupled to gravity. This is naturally thought of as part of the swampland program. We are thus motivated to try to sharpen a general statement along these lines. 

Thus far we have used terms such as ``EFT," ``approximate symmetry," etc. in the usual sense that they are understood in the literature. However, to formulate a more specific conjecture we will wish to restrict our definition of these terms somewhat. We give the following definitions which reflect certain assumptions we will make: 

\begin{itemize}
    \item We consider Lorentz-invariant effective field theories (EFTs)  defined in terms of 
     \begin{itemize}
         \item A specification of the gauge group and field content, with a finite number of fields 
         \item A Lagrangian $\mathcal{L}_\mathrm{EFT}$ constructed out of local operators
         \item A cutoff $\Lambda$ indicating the maximum energy scale at which the EFT is valid (\textit{i.e.} fields with mass above $\Lambda$ are omitted in the EFT description).
     \end{itemize}
     Note that a full quantum gravity model such as string theory is generally not described by an EFT involving a finite number of local fields. EFTs are limited to providing low-energy descriptions of a full theory.
    \item We say that such an EFT has an exact linearly realized continuous global symmetry $G$ if there is some infinitesimal linear transformation  of the fields which leaves the path integral invariant and which is \emph{not} a subgroup of the gauge transformations of the Lagrangian. (We will generally shorten ``linearly realized continuous global symmetry" to simply ``global symmetry" or even just ``symmetry.")
    \item We say that an EFT has $G$ as an approximate symmetry if setting the coefficients of some set of operators in the Lagrangian produces an EFT with $G$ as an exact symmetry. 
\end{itemize}

We can now express in a more precise way the expectation that within quantum gravity, local processes described by EFT violate global symmetries faster than black holes. First let us provide a definition:
 \\
 
\fbox{\begin{minipage}{14cm}
{\it
{\bf Local rate bound (definition)}: An EFT with cutoff $\Lambda$ which has an approximate global symmetry $G$ is said to satisfy the \emph{local rate bound} if for any temperature $T<\Lambda/8\pi$ the rates of $G$ violation by thermal black holes $\Gamma^{\not G}_{\rm BH}$ and by local processes $\Gamma^{\not G}_{\rm EFT}$ satisfy the approximate inequality
 }
 \begin{equation} \quad\quad\quad\quad\quad\quad\quad\quad \Gamma^{\not G}_{\rm BH}\lesssim \Gamma^{\not G}_{\rm EFT}
\,.  \label{eq:TRB} ~~~~~~~~~~~~~~~~~~~~~~~~
 \end{equation}
\end{minipage}
}
\vspace{5pt}
\\
\newcommand{\Lmax}{\Lambda_\mathrm{EFT,max}}
The implications of \Eq{eq:TRB} for the form of symmetry-violating EFT operators will be discussed in \Sec{sec:TRBoperator}. 

Our discussion in Sec.~\ref{se:thermal} motivates the hypothesis that for any quantum gravity theory the bound \Eq{eq:TRB} is satisfied by the ``sub-Planckian  EFT", meaning the most complete possible EFT description with the highest possible cutoff. Starting from this EFT, integrating out modes produces lower-energy EFTs which describe the dynamics accessible to a lower-energy observer. When exploring emergent symmetries in QFT in \Sec{se:Models}, we will often find that if a QFT satisfies the local rate bound \Eq{eq:TRB}, then low-energy EFT descriptions also satisfy it. 

However it is possible that as fields are integrated out from the sub-Planckian EFT, the action of the global symmetry on the remaining fields becomes exact enough to violate \Eq{eq:TRB}. One example in which an extremely exact global symmetry is essentially forced in a low-energy EFT by gauge symmetry will be discussed in \Sec{sec:multi}; another occurs in a ``clockwork" model with a large discrete symmetry (\Sec{sec:clock}).

In these examples, while the low-energy EFT has a very exact global symmetry, it is always completed below the Planck scale into an EFT in which the global symmetry is violated at a level consistent with the local rate bound \Eq{eq:TRB}. This suggests that the local rate bound is indeed generically satisfied within some EFT description of the theory, though not necessarily at arbitrarily low energy. In light of this, we propose the following statement as a possible general constraint on EFTs descending from a quantum gravity theory:
\\

\fbox{\begin{minipage}{14cm}
{
\textbf{Swampland Symmetry Conjecture (SSC)}: \it In the presence of gravity, any effective field theory with exact or approximate global symmetries is UV-completed into an effective field theory with some cutoff $\Lambda \lesssim \mPl$ that has no exact global symmetries and for which all approximate symmetries satisfy the local rate bound \Eq{eq:TRB}. 
}
\end{minipage}
}
\\

Some comments are in order. The various EFTs discussed here should be understood as descriptions of the same underlying quantum gravity theory, just with different Wilsonian energy cutoffs. ``UV-completion" here refers to raising the Wilsonian cutoff of the EFT description, not to completion into a (non-EFT) quantum gravity model. In the case that the original EFT itself satisfies the local rate bound, the conjecture is satisfied without considering a higher-energy completion.

The conjectured local rate bound relies on considerations on Boltzmann suppression, and as such is independent of spacetime dimension. Hence the conjecture is not restricted to 4D spacetime (though in this case the ``Planck scale" should be understood as the appropriate $d$-dimensional gravity scale). UV-completion of an EFT may involve increasing the spacetime dimensionality, for example when the cutoff is raised above the size of a compact extra dimension.
In such case, the extension of local Lorentz invariance to higher dimensions   should be taken into account in SSC considerations, since Lorentz-invariance of the EFT is a  condition required in our formulation of the SSC.

Note that in principle the SSC defined above does not directly constrain any EFT with a sub-Planckian cutoff---one could always imagine that the local rate bound is satisfied by some additional physics appearing just below the Planck scale. However, given the assumption of the conjecture that some such physics is necessary, it does not seem generic for it to be ``postponed" to the highest possible scale. One may expect that it is more typical that an EFT description well below the Planck scale satisfies the local rate bound. The situation may be compared to that of the Weak Gravity Conjecture (WGC). The most minimal forms of the WGC (see \Sec{sec:wgcreview}) could be satisfied by new particles at the Planck scale, and thus in principle do not constrain sub-Planckian physics. However, in specific examples the physics responsible for guaranteeing the WGC appears at parametrically sub-Planckian scales and has an impact on EFT dynamics. Similarly we will find (\Secs{se:WGC}{se:Models}) that the local rate bound \Eq{eq:TRB} is typically satisfied well below the Planck scale, and often holds even in very low-energy EFTs where most particles have been integrated out.

\subsection{Sufficient bounds on a global symmetry violating operator}
\label{sec:TRBoperator}

Let us return to the local rate bound \Eq{eq:TRB} and consider how an EFT may satisfy it. Note that the inequality \Eq{eq:TRB} is trivially satisfied at $T=0$, for which the rates on both sides vanish. At nonzero temperature, the local rate bound expresses a lower bound on the $G$-violating local operators.

In the case where a single operator ${\cal O}_{\not G}$ dominates the EFT rate,  simple bounds on ${\cal O}_{\not G}$ follow from the local rate bound. Let us consider the important case of a single 4D operator with a moderate amount of derivatives, such that the number of fields in the operator is approximately the dimension of the operator, $d_{\not G}\approx n_{\not G} $. For a given such operator 
\be
{\cal L}_{\not G}\supset \frac{c_{\not G}}{\Lambda^{d_{\not G}-4}} {\cal O}_{\not G}\,,
\ee
the conjectured bound of Eq.~\eqref{eq:TRB} is satisfied if the following \textit{sufficient} conditions are satisfied:
\\

 \fbox{\begin{minipage}{14cm}
 \begin{align}
&{\rm Either~of}   &\log c_{\not G} \gtrsim - \frac{4\pi}{ G_{\rm N} \Lambda^2}\quad\quad\quad  {  (log-coefficient~bound)}   \label{eq:logcbound} \\
&   \quad &d_{\not G} \log d_{\not G} \lesssim  \frac{4\pi}{ G_{\rm N} \Lambda^2}
\quad\quad\quad\quad\quad    {  (dimension~bound)} \label{eq:dimbound}
  \, \\
 &{\rm And}   &\sum_{i=1}^{n_{\not G}} m_i
\lesssim \frac{1}{G_{\rm N}\Lambda}\,  
\quad\quad\quad\quad\quad\quad\quad\quad~   {  (mass~bound)} \label{eq:massbound}
\end{align}
 \end{minipage}
 }
\\

So far we have focused on 4D theories, but the local rate bound \Eq{eq:TRB} and the SSC can readily be applied to higher-dimensional models, which in fact provide interesting realizations of approximate symmetries (\Sec{sec:flat}). The analysis of \Sec{se:bound} proceeds essentially unchanged except that the mass-radius relationship for $D$-dimensional black holes is $\MBH \sim G_{\mathrm{N},D} \RBH^{D-3}$ where $G_{\mathrm{N},D}$ is the $D$-dimensional Newton's constant. As before we can identify the minimal black hole radius with a cutoff scale $\Lambda^{-1}$ to obtain bounds analogous to the above. For example, the ``mass bound" \Eq{eq:massbound} would now have the mass of the minimal $D$-dimensional black hole on the right hand side.

\section{Accidental symmetry from gauge theory and the WGC}
\label{se:WGC}

The arguments against global symmetries in gravity suggest that one should not impose such symmetries to forbid certain operators in an EFT Lagrangian. However, as discussed above the gauge structure of a model may enforce an ``accidental" global symmetry at low energies by forbidding all low-dimension operators that would violate it. In principle one may imagine writing EFTs in which the most relevant operator that can violate a global symmetry has very high dimension. For example let us consider an EFT with a $U(1)$ gauge group with coupling constant $g$ and two complex scalars $\psi, \chi$ with (integer) $U(1)$ charges $q_\psi$, $q_\chi$. Let us denote the particle numbers (\textit{i.e.} number of particles minus number of antiparticles) for the two fields as $N_\psi$, $N_\chi$. Then $q_\psi N_\psi + q_\chi N_\chi$ corresponds to the exactly conserved gauge charge, but the individual particle numbers $N_\psi$, $N_\chi$ may also be conserved up to the effects of high-dimension operators, depending on the charges $q_\psi$, $q_\chi$. For example, if $q_\psi$, $q_\chi$ are mutually prime, then the lowest dimension operator which can violate the global particle numbers is 
\eq{
\label{eq:wgcop}
\frac{c}{\mu^{q_\psi+q\chi-4}} \psi^{q_\chi} \left( \chi^\dagger \right)^{q_\psi} + \mathrm{ h.c.}
}  
This operator dimension $d = q_\psi + q_\chi$ could be taken arbitrarily large if the integer $q_i$'s are chosen to be large. In this case the theory could violate the local rate bound \Eq{eq:TRB}, specifically by violating the ``operator dimension bound'' of \Eq{eq:dimbound} and/or the ``mass bound'' \Eq{eq:massbound} depending on the particle masses. Maintaining perturbative control requires that $q_\psi, q_\chi \lesssim 4\pi/g$, but this still allows for large charges if the coupling $g$ is very small. 

From a pure quantum field theory perspective, nothing is inconsistent about writing the previous model, where we chose a specific charged particle content for a $U(1)$ gauge theory with small coupling $g$. However, the Weak Gravity Conjecture (WGC) in its various forms provides constraints on gauge EFTs which emerge as low-energy descriptions of a quantum gravity theory. In general, it prevents taking the limit of $g \to 0$ in a theory with gravity and places some requirements on the spectrum of charged particles. Below (\Sec{sec:wgcreview}) we will briefly review some of the proposals that go under the label of ``WGC'' before discussing how they bound the rate of global symmetry violation in a way that is reflected by the Swampland Symmetry Conjecture \Sec{sec:wgcglobal}.

\subsection{Review of WGCs}
\label{sec:wgcreview}

There are many distinct statements and proposals that collectively fall under the umbrella of ``Weak Gravity Conjecture(s).'' (For a recent review, see section 3.5 of~\cite{Palti:2019pca}.) The most minimal versions require the existence of charged particles that are light enough to allow decay of extremal charged black holes, \textit{i.e.} black holes with the maximal charge-to-mass ratio without featuring a naked singularity. For a single $U(1)$ gauge factor, a non-rotating extremal black hole with charge $Q g$ has mass $M = Q g \mPl$.\,\footnote{The normalization for the gauge charge $q g$ used here in the context of the WGC is larger by a factor of $\sqrt{2}$ compared to the canonical normalization for gauge couplings, \textit{i.e.} when the gauge kinetic term is written as $\mathcal{L} \supset -\frac{1}{4 g_\mathrm{can}^2} |F|^2$. See \textit{e.g.} Sec. 3.3 of \Ref{Palti:2019pca}. We will ignore this $O(1)$ factor throughout.}  The simplest form of the WGC requires that there exists some charged state that is ``superextremal'', \textit{i.e.} with charge greater than its mass in these units:
\bea
\label{eq:ewgc}
m < q g \mPlr.
\eea
Extremal black holes would be able to decay by emitting such a state, leaving the black hole subextremal ($M \geq Q g \mPlr$), so that  there is no naked singularity. Note however that this condition could be satisfied by a particle with large charge $q$ (as large as $\sim 4\pi/g$ while remaining perturbative) and mass close to the Planck scale, and thus does not necessarily constrain sub-Planckian physics. 

There are much stronger forms of the WGC which propose that there must exist particles satisfying \Eq{eq:ewgc} for many different values of the integer charges $q$, motivated by a variety of considerations. One proposal of this sort is the Lattice WGC~\cite{Heidenreich:2015nta}, which requires that for every possible charge in the charge lattice of a theory (\textit{e.g.} every value of $q_A$ and $q_B$ in a theory $U(1)_A \times U(1)_B$) there exists a superextremal particle; \textit{i.e.}  with mass less than its total charge $\sqrt{\sum_i q_i^2 g_i^2} \mPl$. This requires an infinite tower of states with typical mass splittings $\sim g \mPlr$. This signals the eventual breakdown of local 4D EFT, similar to the case of a KK tower arising from a compact extra dimension. 

However, some counterexamples to the Lattice WGC have been demonstrated. For example, in~\Ref{Heidenreich:2016aqi} an example based on Kaluza-Klein theory was given in which particles of odd charge under a gauge factor are $O(1)$ heavier than the extremality bound. This suggested that perhaps superextremal particles always populate at least a sublattice of the full charge lattice~\cite{Heidenreich:2016aqi}. However even this refined conjecture is violated in the examples of \Ref{Lee:2019tst}---although in these, there always exists a minimally charged particle which was superextremal.   

In light of these examples, a more valid conjecture may be the less specific  ``Tower WGC'' stated in \Ref{Andriolo:2018lvp}, which simply suggests that there must exist a sequence or ``tower'' of charged particles (not necessarily occupying a sublattice) satisfying some bounds on their charge-to-mass ratios. This would suggest that the presence of particles of various charges satisfying the extremality bound \Eq{eq:ewgc} is generic, though perhaps not guaranteed for any particular charge value.\,\footnote{\Ref{Saraswat:2016eaz} discusses another way in which strong conjectures such as the Lattice WGC could fail: a theory satisfying such a conjecture in a Coulomb phase could violate it in a Higgsed phase where only some of the gauge fields are massless. For the purposes of satisfying the SSC, it is sufficient for some completion of the theory (\textit{e.g.} above the Higgsing scale) satisfies the Lattice or Tower WGC.}

In what follows (\Sec{sec:wgcglobal}), we will generally assume that there exist particles with minimal charge under the $U(1)$ gauge factors which satisfy \Eq{eq:ewgc}, which will be sufficient to guarantee that the SSC is satisfied. We will also briefly discuss the possibilities when this sufficient condition does not hold, as has been observed in a few examples.

These considerations can be applied to models with multiple $U(1)$'s, which actually suggest stronger bounds on particle masses than the single $U(1)$ case \Eq{eq:ewgc}. Black holes carrying arbitrary charges $Q_i g _i$ under multiple $U(1)$'s satisfy the extremality bound $M_\mathrm{BH}^2 \geq \sum_i Q_i^2 g_i^2 \mPl^2$. Requiring that any extremal black hole can decay implies a constraint on the charged particle spectrum defined in \Ref{Cheung:2014vva} (the ``convex hull condition''). For illustrative purposes let us consider the simplifying case of $N$ identical $U(1)$'s, \textit{i.e.} with common gauge coupling strength $g$, and suppose that there exist unit-charged particles under each $U(1)$ with common mass $m$. Then the convex hull condition is satisfied if
\eq{
\label{eq:multiwgc}
m \leq g \mPl/\sqrt{N}.
} 

\subsection{WGC and global symmetry violation}
\label{sec:wgcglobal}

In light of the WGC let us revisit models of the type discussed at the start of \Sec{se:WGC}, in which a low-energy $U(1)$ gauge EFT contained a particle $\psi$ with large charge $q_\psi$ such that global $\psi$-number could only be violated by very high-dimension operators. If we assume the Lattice WGC, then any such EFT must eventually be completed into a theory containing a tower of superextremal particles with every possible $U(1)$ charge, \textit{i.e.} with mass spacing $\lesssim g \mPl$. For simplicity let us consider extending the EFT description to include only the first particle of this tower, denoted $\chi$, with mass $\leq g \mPl$. Since this particular EFT does not describe the rest of the tower, it has a cutoff $\Lambda \lesssim 2 g \mPl$ (otherwise we are obligated to consider the heavier states as well). Within this EFT we may write an operator violating $\psi$-number
\eq{
\frac{c}{\Lambda^{q_\psi - 3}} \psi^\dagger \chi^{q_\psi}  + \mathrm{ h.c.}
\label{eq:psiop}
}
Consistent with the idea that $\psi$-number is a purely accidental or emergent symmetry, we will assume that the coefficient of this operator is unsuppressed (\textit{i.e.} $c$ is not exponentially small). However, its dimension $d \sim q_\psi$ could be quite large. In order to accurately compute interaction rates we must assume that the $U(1)$ theory is perturbative, so that $g q_\psi \lesssim 4\pi$. This places an upper bound on the dimension of this operator:
\eq{
\label{eq:WGCdbound}
d \lesssim 4\pi/g\,.
}
We can compare this to the ``dimension bound'' of \Eq{eq:dimbound}, $d \log d \lesssim \mPl^2/\Lambda^2$. Using $\Lambda \sim g \mPl$, this is equivalent to $d \log d \lesssim 32 \pi^2/g^2$. We see that for small $g$ the condition \Eq{eq:WGCdbound} guarantees that the conjectured dimension bound is satisfied.

We can also consider the ``mass bound'' \Eq{eq:massbound}. Since most of the particles produced in the global symmetry violating interaction are $\chi$'s, the bound reads $d m_\chi \lesssim (G_N \Lambda)^{-1} \sim 8 \pi \mPl/g$. The WGC constraints meanwhile give 
\eq{
d m_\chi \lesssim \left(\frac{4\pi}{g}\right) \left(g \mPl \right) = 4\pi \mPl.
}
Thus the mass bound \Eq{eq:massbound} is parametrically satisfied for small $g$, and saturates for $g \sim 1$ (for which $\Lambda \sim \mPl$, at the limit of theoretical control). 

Thus far we have considered an EFT description with cutoff just above the WGC scale $g \mPl$, which we have assumed contains a unit-charged superextremal particle. If one considers extending the EFT to higher energies, then the Lattice WGC would indicate that additional superextremal particles will appear, of every possible charge. Using these higher-charged particles one could write operators of lower dimension which violate $\psi$-number. However, if we are considering a higher cutoff for an EFT we should allow for smaller black holes contributing to the thermal rate, reflected in the right hand sides of \Eqss{eq:logcbound}{eq:dimbound}{eq:massbound} decreasing for larger $\Lambda$. Thus both sides of the local rate bound vary as the cutoff is increased. 

Therefore we should check that these bounds are still satisfied if we take an EFT cutoff of $\Lambda \sim Q g \mPl$ for any integer $Q$, and allow for superextremal particles with charge up to $Q$. Using particles of charge up to $Q$ coud allow operators of dimension $d \sim q_\psi/Q$ to violate the global symmetry, but as these could have mass as great $Q g \mPl$, the total mass of the particles required in the operator remains $\sim q_\psi g \mPl$, approximately independent of $Q$. Again, this total mass is bounded as $4\pi \mPl$ in a perturbative theory, which should be compared to the minimal black hole mass $8\pi \mPl^2/\Lambda$. Thus the bound is still satisfied for any $\Lambda \lesssim \mPl$. Similar reasoning shows that the operator dimension bound \Eq{eq:dimbound} is also satisfied. 

Thus far we have taken the Lattice WGC as an assumption and shown that it is a sufficient condition for the Swampland Symmetry Conjecture to be satisfied. However, as discussed in \Sec{sec:wgcreview} there are examples of UV completions of gauge theories in which the Lattice WGC is not satisfied, \textit{i.e.} there does not exist a superextremal particle for every possible charge. The ``Tower WGC'' however suggests that there is some infinite sequence of superextremal states in the charge lattice, even if not all lattice sites are ``occupied.'' There are many possibilities for such a sequence to satisfy the SSC. As discussed, one could use predominantly higher charge states to induce a symmetry violating process at a rate consistent with the SSC. A serious obstacle to satisfying the necessary bounds would occur if there were not enough different charged states in the theory to cancel out the charge $q_\psi$ without requiring very large total mass. For instance, in the example considered at the beginning of \Sec{sec:wgcglobal}, the only charged particles had charges $q_\psi$, $q_\chi$. If these are coprime and large, then it is possible to violate the mass bound even if both particles are superextremal. It seems likely that there could exist some formulation of a Tower WGC which is weaker than the Lattice WGC (\textit{i.e.} consistent with the counterexamples to the latter) but strong enough to rule out such scenarios. 

\subsubsection{Higher-dimensional gauge theory} 
\label{sec:5DWGC}

Another interesting case to check is a scenario with a compact extra dimension which appears below the WGC scale $\sim g \mPl$. For example, for a flat extra dimension of size $L$, the 5D quantum gravity scale is $M_5^3 \sim \mPl^2/L$, and 5D black holes could exist with masses as small as $\sim M_5$. Allowing for such small black holes would alter \Eqss{eq:logcbound}{eq:dimbound}{eq:massbound} to give tighter bounds, as discussed in \Sec{sec:TRBoperator}. One may be concerned that this would lead to violation of the local rate bound and the SSC.

To check this let us return to the example considered above in which an approximate symmetry appeared due to a particle $\psi$ with large integer charge $q_\psi$. However, let us now assume that above some scale $1/L$ this model is completed into a 5D gauge theory. The 5D gauge coupling $g_5$ is non-renormalizable and is related to the 4D coupling $g$ by $g_5^2 \sim g^2 L$. The WGC in 5D (see \textit{e.g.} \Ref{ArkaniHamed:2006dz, Palti:2019pca}) requires particles with mass $m < q g_5 M_5^{3/2}$, which is in fact equivalent to the 4D condition $m < q g \mPl$. 

However, since the $U(1)$ theory is now non-renormalizable, EFT control necessarily breaks down at a scale $\Lambda_5 \sim (q_\psi g_5)^{-2}$. Thus we can only consider black holes of radius larger than $\Lambda_5^{-1}$, or mass greater than $\MBHm \sim M_5^3 \Lambda_5^{-2}$, while retaining theoretical control (assuming $\Lambda_5 < M_5$). Recall that the dimension of the global symmetry violating operator \Eq{eq:psiop} is constrained as $d \lesssim q_\psi$, and the total mass of the states involved in this operator is $\sim d m_\chi \sim q_\psi g \mPl \sim \left(\Lambda_5 L\right)^{-1/2} \mPl$. Using $\mPl L^{-1/2} \sim M_5^{3/2}$, we see that this mass is always less than the mass of the minimal 5D black hole $\MBHm$, so that the mass bound \Eq{eq:massbound} is still satisfied. One can similarly check the operator dimension bound.

\subsubsection{Multiple $U(1)$'s}
\label{sec:multi}

One could make considerations similar to the above for theories with multiple $U(1)$ gauge symmetries. Let us first consider the extent to which accidental global symmetries can appear in models with multiple $U(1)$'s, before considering how multi-field versions of the WGC could ensure that the SSC is ultimately not violated.

Generalizing the scenario discussed at the start of  \Sec{se:WGC}, one can consider a theory with $N$ different $U(1)$ gauge fields and $N$ charged particles, $\chi_1, \ldots \chi_N$, such that the charge vectors of the particles span the space of all possible charges. Now introduce another charged particle $\psi$. $\psi$-number will not be exactly conserved, since by assumption its gauge charge
can also be carried by some collection of $\chi$ particles. Thus there can exist some operator involving $\psi$ and the $\chi_i$ which violates $\psi$-number. However, such an operator can again have very large dimension depending on the specific charges involved.

An extreme example is obtained by considering a ``gauge clockwork" model as in~\cite{Saraswat:2016eaz}, where we take the $\chi$'s to have charges $\mathbf{q}_{\chi_1} = (1, -Z, 0, \ldots, 0)$, $\mathbf{q}_{\chi_2} = (0, 1, -Z, 0, \ldots, 0)$ etc., up to $\mathbf{q}_{\chi_N} = (0, 0, \ldots, 1)$. Then if we take $\psi$ to have charge $(1,0,\ldots,0)$, the lowest dimension operator violating $\psi$-number will be $\psi^\dagger \chi_1 \chi_2^Z \chi_3^{Z^2} \ldots \chi_N^{Z^{N-1}}$, \textit{i.e.} with exponentially large dimension. This can greatly violate the operator dimension bound of \Eq{eq:dimbound}, even if the gauge couplings $g_i$ of the $U(1)$'s are \emph{not} taken to be small! (As before, perturbativity merely requires $Z \lesssim 4\pi/g_i$.) Since the gauge couplings could be $O(1)$, even strong versions of the WGC such as the Lattice WGC will not require us to introduce additional charged particles with mass much less than $\mPl$, \textit{i.e.} the cutoff of the ``gauge clockwork" EFT can be close to the Planck scale. This is therefore an example of a low-energy EFT that violates the local rate bound \Eq{eq:TRB}. 

Our statement of the Swampland Symmetry Conjecture allows for such a possibility, provided that the EFT is completed at some scale into a theory that \emph{does} satisfy the local rate bound. We can see that the Lattice WGC would in fact ensure this. Even if the light fields in the model have charges of the gauge clockwork form, the Lattice WGC requires that there exist a superextremal particle at every site in the charge lattice, so that in the complete theory there are many possibilities to create a low-dimension operator violating $\psi$-number. Let us now generalize beyond the gauge clockwork model to consider a particle $\psi$ with charges $q_i$ under various $U(1)$'s. Unless all of the $q_i$ are small, we can always write such an operator violating $\psi$-number using only superextremal particles with smaller total charge than $\psi$, meaning that they are lighter than $\sqrt{\sum_i g_i^2 q_i^2} \mPl \lesssim 4 \pi \mPl$ (the latter inequality following from perturbativity of $\psi$). Thus there is no possibility of parametric violation of the mass bound \Eq{eq:massbound} or the operator dimension bound \Eq{eq:dimbound}.

We again see that the Lattice WGC, with its infinite sequence of charged particles, suffices to enforce the SSC. However, since it is observed that the Lattice WGC is not always satisfied, one would again ideally want to be able to formulate a weaker condition that would still ensure that the SSC is satisfied. Even without considering the most general possible case, it is interesting to ask what types of more minimal charged spectra can guarantee the SSC. 

To this end let us consider the simple model discussed at the end of \Sec{sec:wgcreview}, in which there are $N$ identical $U(1)$ sectors and the ``convex hull" condition of \Ref{Cheung:2014vva} is satisfied by particles $\chi_i$ each with unit charge under one of the $U(1)$'s and with mass $m < g \mPl/\sqrt{N}$. Suppose that in addition to these particles we had some $\psi$ with charges $q_i$ under the various $U(1)$'s. Then the lowest dimension operator we can write which violates $\psi$-number is $\psi^\dagger \prod_i \chi_i^{q_i}$. This has dimension $d \sim \sum_i q_i$. Requiring perturbativity, $\sum_i g^2 q_i^2 \lesssim 16\pi^2$, we have $d \sim \sum_i q_i \lesssim 4\pi \sqrt{N}/g$ (compare to \Eq{eq:WGCdbound}). Using $m_\chi \lesssim g \mPl/\sqrt{N}$, we observe that both the operator dimension and mass bounds are again parametrically satisfied (and potentially saturated for large $g$). 

Thus we see that we do not necessarily require a densely populated lattice of charges in order to ensure that the SSC is satisfied. As with the single-field case, it is interesting to consider whether a more precise formulation of a multifield ``Tower" WGC can be made and what implications it would have for the SSC. 

\subsection{Discrete symmetry}
\label{sec:discrete}

Thus far in this section we have considered how continuous gauge symmetry can enforce nearly-exact continuous global symmetries. However, one could also consider theories with exact discrete symmetries, either global or gauged, which could also produce accidentally exact continuous symmetries. The standard argument against global symmetries in gravity based on the finiteness of black hole entropy (\Sec{se:intro}) does not immediately apply to discrete symmetries, since discrete charge can only take on a finite number of values. In the context of models with CFT duals, \Refs{Harlow:2018tng, Harlow:2018jwu} demonstrate that exact global symmetries are inconsistent even if they are discrete, though gauged discrete symmetries remain possible. 

If one allows models with large discrete symmetries and arbitrary charge assignments one can easily arrange for violation of the SSC. For example, if we consider a field $\psi$ that transforms under a $\mathbb{Z}_N$ discrete symmetry, $\psi \to \exp(2\pi i/N) \psi$, and assume no other charged fields, then $\psi$-number can only be violated by operators $\propto \psi^N$, which would violate the SSC if $N \gtrsim 4\pi/\left(\GN \Lambda^2\right)$.

In analogy with the previous discussion, one may expect that there could exist ``swampland constraints" on the size of discrete symmetries or the spectra of fields charged under them, which could forbid scenarios in which the SSC is violated. For example, \Ref{Dvali:2008tq} considers the case of the previous example (a single field charged under a $\mathbb{Z}_N$) and gives the following argument for the bound $N \lesssim 4\pi/\left(\GN \Lambda^2\right)$, which would enforce the SSC (see also a brief discussion in \Ref{Dvali:2007hz}). They consider black holes of size close to the cutoff $\Lambda^{-1}$, with mass $\MBH \sim (\GN \Lambda)^{-1}$ and temperature $\TBH \sim \Lambda/4\pi$. If these have the maximal $\mathbb{Z}_N$ charge $\sim N$, and lose charge only through thermal radiation of $\psi$'s, with typical energy $\TBH$, they can only evaporate without leaving any remnant if $\MBH \gtrsim N \TBH$, leading to the above bound on $N$. However, this argument is not conclusive, as it is possible that charge could be lost through non-thermal processes. A weaker form of the same argument thus requires simply $\MBH \gtrsim N m_\psi$~\cite{Dvali:2008tq}, which at least ensures that the ``mass bound" condition \Eq{eq:massbound} is satisfied. 

If one generalizes the conclusions of \Refs{Harlow:2018tng, Harlow:2018jwu} beyond the AdS/CFT context, it would suggest that exact discrete symmetries should always be gauged. This may allow for additional arguments constraining the nature and size of the symmetry. A familiar case in which discrete gauge symmetries appear is the Higgsing of an Abelian gauge field by a particle of non-minimal charge. If we assume this UV completion for a discrete gauge symmetry then one can apply the WGC to constrain the cutoff of the EFT just as we have discussed in \Sec{sec:wgcglobal}. \Ref{Craig:2018yvw} proposes some constraints on discrete gauge symmetries without such assumptions, though as yet there is no statement comparable to that of the Tower WGC for continuous gauge groups. 

In summary, making a complete determination of whether or not accidental continuous symmetries could violate the SSC requires further understanding of the ``swampland'' of exact discrete symmetries. 

\section{SSC in QFT models of  emergent symmetries }
\label{se:Models}

A variety of quantum-field theoretical models may be used to obtain near-exact global symmetries. Such models/mechanisms are ubiquitous in particle physics model-building, including for instance  models giving rise to hierarchical Yukawa matrices rendering approximate a part of the SM global symmetries. Without gravity, such models of emergent approximate symmetries can be pushed to the extreme, generating arbitrarily small symmetry-violating operators, and thus giving rise to exact global symmetries.

In contrast, in the presence of gravity,  obstructions to such limit are expected to occur, motivated at the qualitative level from general quantum gravity considerations and at the quantitative level from the SSC. When confronting a given QFT model to the SSC, 
 a possibility is left open  that the model under consideration does not satisfy the local rate bound, and that instead only its UV completion in a higher energy EFT---whose existence is conjectured by the SSC---satisfies the bound (see Sec.~\ref{se:swamp}).
However, it is certainly interesting to verify whether or not the local rate bound  is satisfied by the QFT model itself. When this is true, it readily implies that the SSC is valid without invoking the possibility of an embedding into a more fundamental EFT.

QFT models for emergent symmetries often have various mass scales, and can thus reduce to a variety of different EFT with lower cutoff. In such cases it is interesting to study the various energy regimes, considering systematically what observers with access to specific energies would conclude when evaluating the local rate bound.
Studying these models also gives  the opportunity to
illustrate the fate of the emergent symmetry at high energies, as perceived by the low-energy observer. 

\subsection{The Frogatt-Nielsen mechanism}

The Frogatt-Nielsen (FN) mechanism~\cite{Froggatt:1978nt} was originally proposed to explain the small and hierarchical couplings for the Yukawa operators of the Standard Model, schematically written as $\bar \Psi \Psi' H$. The basic idea is to assume the presence of a higher-dimensional operator such as 
\be
{\cal L}_{\rm FN}\supset c\left(\frac{\phi}{M}\right)^n \bar \Psi \Psi' H \,\,+\,\,{\rm h.c.}\, \label{eq:FN_start}
\ee
in the Lagrangian, where $c=O(1)$, $\phi$ is a complex ``flavon'' field and the others field  mimic a Standard Model Yukawa operator.\,\footnote{The $c$ coefficient could be taken as small as the log-coefficient bound, however since our bounds are logarithmic and approximate, nothing is gained conceptually by taking a small $c$ and it is enough to take $c=O(1)$.  }
 This operator violates $\Psi$-number, which is thus  an approximate global symmetry. 
 An exact $U(1)_F$  symmetry is \textit{assumed}  in order to forbid operators of lower order in Eq.~\eqref{eq:FN_start}, with the $U(1)_F$ charges satisfying $q_{\Psi}+q_{\Psi'}+q_H+n q_\phi=0$. This $U(1)_F$ may be either global or gauged, which will crucially change the subsequent reasoning. 
  One then assumes that the $\phi$ field spontaneously breaks the $U(1)_F$, with $\langle\phi\rangle=v$. In the  broken phase, $\Psi$-number is now violated by the dimension-four operator  
\be
{\cal L}\supset c_{\rm b} \bar \Psi \Psi' H \,\,+\quad{\rm h.c.}\,,\quad c_{\rm b} = c \left(\frac{v}{\Lambda}\right)^n \label{eq:FN_LE} \,
\ee
where the operator coefficient $c_{\rm b}$ can be exponentially small for large $n$, hence giving rise to  an arbitrarily small violation of $\Psi$-number. Let us now analyze this mechanism from the perspective of the SSC.

Let us first briefly discuss the case of ungauged $U(1)_F$.  
In this case the starting point ${\cal L}_{\rm FN}$ is a QFT with an exact global symmetry, which is of course  inconsistent with our conjecture, unless ${\cal L}_{\rm FN}$ is UV-completed by an EFT in which $U(1)_F$ is approximate or absent. Note that in the language of the SSC, such UV completion may simply consist in appending  $U(1)_F$-violating operators to ${\cal L}_{\rm FN}$. In such case $\Psi$-number-violating operators of lower order such as  $\bar \Psi \Psi' H$  are allowed, such that the FN mechanism is spoiled.

More interestingly, let us assume that the $U(1)_F$ is gauged, with coupling constant $g_F$.
In that case ${\cal L}_{\rm FN}$ has no exact global symmetry. Since ${\cal L}_{\rm FN}$ is a gauge theory, it  is constrained by the  weak gravity conjecture(s), exactly as in the scenario discussed in \Sec{se:WGC}.  
 Requiring large $n$ means requiring large $|(q_{\Psi}+q_{\Psi'}+q_H)/q_\phi|$.
 Let us take the most extreme case where $q_\phi=1$, and take large $q_\Psi>0$ for concreteness, such that $n \sim q_\Psi$. 
 For the EFT to remain weakly coupled, we should also require $g_F q_\Psi \leq 4\pi$.
One may notice that taking $n$ to infinity would mean taking  $g_F\rightarrow 0$, which is prohibited by the WGC. Let us thus keep a finite, large $n$. The Tower WGC (see \Sec{sec:wgcreview}) indicates that an EFT containing only one particle with $O(1)$ charge can be a valid description only up to the scale $\Lambda \sim g_F \mPlr$. We thus have
\be
 n \lesssim 4\pi \mPlr/\Lambda \label{eq:FN_WGC}\, 
 \ee
as in \Eq{eq:WGCdbound}, which ensures that the dimension bound \Eq{eq:dimbound} implied by the SSC is satisfied. As in \Sec{sec:wgcglobal} we may also consider the mass bound \Eq{eq:massbound} and find that it is potentially saturated, but not violated.

Interestingly, whether one considers the  theory in the Higgsed or unbroken phase,  the approximate bound from the SSC remains the same. In the broken phase, the relevant bound is the one on the log-coefficient Eq.~\eqref{eq:logcbound}, giving roughly $\log c_{\rm b} \gtrsim - 4\pi/G_{\rm N}\Lambda^2$. In the unbroken phase, in the limit of large dimension which is the one considered here, the relevant bound is rather the dimension bound Eq.~\eqref{eq:dimbound}, roughly $n \lesssim  4\pi/G_{\rm N}\Lambda^2$. Since $\log c_{\rm b} \sim - n$, it turns out the two bounds are  approximately equivalent. Such feature is specific to the FN mechanism and was no guaranteed a priori.

Physically, how does the WGC enforce a large enough violation of $\Psi$-number? Though we have only included particles with specific large charges in ${\cal L}_{\rm FN}$, the Tower WGC indicates that some particles $\tilde{\phi}$ with $O(1)$ charges should also exist. Other operators of lower dimension such as $\bar \Psi \tilde \Psi' H (\tilde \phi/M)^{\tilde n}$, $\tilde n \ll n$, can be formed, such as the local rate bound on $\Psi$-number violation is satisfied.

For a low-energy observer knowing only about $c_{\rm b}$, what does using the SSC bound implies? Using the log-coefficient bound Eq.~\eqref{eq:logcbound} on the measured value of $c_b$, the observer concludes that the local  rate bound is violated  when the a priori-unknown cutoff scale is taken above the value $\Lambda_{\rm SSC}$ given by $\Lambda_{\rm SSC} \equiv \mPlr \sqrt{-32\pi^2/\log c_{\rm b}}$.  For small enough $c_{\rm b}$,  $\Lambda_{\rm SSC}$ can  be below $ \mPlr $. 
From this fact, the low-energy observer can suspect that new  physics below the $\Lambda_{\rm SSC}$ scale exists.\,\footnote{The observer cannot conclude for sure that new physics exists below $\mPlr$ because the SSC always leaves open the possibility that the local rate bound is only satisfied in the more fundamental EFT, which may be valid up to the Planck scale. }
Increasing the energy scale in the  experiment, the low-energy observer ends up discovering the  gauge bosons with mass $\sim g_F v$, and at a higher scale the new particles with mass 
$\sim g_F \mPlr$.\,\footnote{ Optionally, the observer may have  predicted this new mass scale after discovering the theory is gauged, by using the WGC. } Hence the hint provided by the local rate bound proved to be fruitful: There was indeed new physics below the $\Lambda_{\rm SSC}$ scale. 
\subsection{The clockwork mechanism}

\label{sec:clock}

The ``clockwork'' mechanism~\cite{Dvali:2007hz, Dvali:2008tq, Choi:2014rja, Higaki:2014pja, Choi:2015fiu, Kaplan:2015fuy}  allows increased control over global symmetry breaking in QFT. These proposals were originally motivated by the goal of achieving a large field space for an axion (as required for models of cosmic inflation, cosmological relaxation~\cite{Graham:2015cka}, etc.), which involves considering an approximate symmetry in a non-linearly realized phase. However the mechanism can also be thought of as a way of producing very exact linearly realized symmetries~\cite{Dvali:2007hz, Dvali:2008tq}, which is the scenario we will focus on. 

\newcommand{\Uc}{U(1)_\mathrm{clock}}

Let us consider a clockwork model as in \Ref{Dvali:2007hz, Dvali:2008tq, Kaplan:2015fuy}, featuring $N+1$ complex scalar fields $\phi_k$, $k = 0,\ldots N$. First let us assume that each $\phi_k$ has a potential invariant under a $U(1)$ global symmetry which gives it a mass. Then add interactions between the $\phi_k$ in the following pattern:
\eq{
\label{eq:clock}
\mathcal{L}_\mathrm{int} \supset &\sum_{k=0}^{N-1} \kappa_{k+1} \phi_k^3 \phi_{k+1}^\dagger + \mathrm{h.c.} \\
= &\kappa_1 \phi_0^3 \phi_1^\dagger + \kappa_2 \phi_1^3 \phi_2^\dagger + \ldots + \kappa_N \phi_{N-1}^3 \phi_N^\dagger + \mathrm{h.c.}
} 
The individual $U(1)$ symmetries are now broken to a single $\Uc$ in which each $\phi_k$ transforms with charge $q_i = 3^k$. There are a variety of ways to break this remaining $\Uc$ global symmetry. For example if we add a linear operator $\mu^3 \phi_M$ for some $M$, then the $U(1)$ is broken down to a discrete $\mathbb{Z}_{3^M}$ subgroup, in the process of which all of the $\phi_k$ with $k \geq M$ obtain vevs. Intuitively, we may expect that since the interaction Lagrangian \Eq{eq:clock} is ``local'' in the $\phi_k$, the fields with $k$ much different from $M$ will feel suppressed violations of the continuous $\Uc$. We will consider two different limits to illustrate this effect and relate it to the SSC.

First, let us consider the case in which we choose to write $\mu^3 \phi_0$ as the explicit symmetry violating operator. This completely breaks $\Uc$ leaving no discrete subgroups. One could of course consider writing additional symmetry violating operators, but they would not break any further exact symmetries, and they are not needed to satisfy the SSC at high energies where all the $\phi_k$ are accessible in the thermal bath. However, let us consider the case in which $\phi_N$ is much lighter than the other $\phi_k$, so that there is a low-energy EFT in which only $\phi_N$ is present. For simplicity let us take all the other $\phi_k$ to have a common mass $M$ and take all the $\kappa_k$ to be equal. Then from writing $\mu^3 \phi_0$ all of the $\phi_k$ with $k < N$ will obtain vevs of order 
\eq{
\label{eq:phivev}
\langle \phi_k \rangle \sim \left(\frac{\mu^3}{M^2}\right) \left(\frac{\kappa \mu^6}{M^6}\right)^{k} \,.
}
Therefore fields with larger $k$ feel exponentially suppressed effects from the symmetry violating operator $\mu^3 \phi_0$. If the field $\phi_N$ has mass $m_N^2$, then in dimensionless terms it feels symmetry violation at the level
\eq{
c \sim \langle \phi_N \rangle^2/m_N^2 \sim \left(\frac{\mu^6}{M^4 m_N^2}\right) \left(\frac{\kappa \mu^6}{M^6}\right)^{2N} \,.
}
Even if $\mu$ is not much less than $M$, indicating that the global symmetry is not very exact in the UV theory, the low-energy EFT of $\phi_N$ can have an exponentially exact symmetry, with $\log c \sim-  N$. 

In pure QFT terms there is no upper bound on $N$, so it seems that the local rate bound (specifically the coefficient bound \Eq{eq:logcbound}) can be violated in this EFT. However, this possibility is avoided if one invokes the well-known conjecture that a large number of particle species necessarily implies a hierarchy between the EFT cutoff and the Planck scale~\cite{Dvali:2001gx, Veneziano:2001ah, Dvali:2007hz, Dvali:2007wp}, 
\eq{
\label{eq:sqrtN}
\Lambda \lesssim \mPl/\sqrt{N}.
}
This implies
\eq{
\log c \sim -N \gtrsim - \mPl^2/\Lambda^2
\label{eq:cclock}
}
satisfying (and potentially parametrically saturating) the coefficient bound \Eq{eq:logcbound}. Just as occurred in \Sec{se:WGC} with the WGC, we find that a previous ``swampland" conjecture, the species bound \Eq{eq:sqrtN}, can enforce the SSC within this model.

Note that \Eq{eq:cclock} is not a \emph{necessary} condition for the SSC---as we have stated it, a low energy EFT may violate the rate bound provided that it is eventually completed into an EFT that satisfies it (in this case, the theory with all of the $\phi_k$ present). It is perhaps suggestive though that even the clockwork mechanism is not powerful enough to produce an EFT that violates the local rate bound when starting from a theory with no exact symmetries. 

Note that if we had chosen the opposite extreme case of writing a symmetry violating operator $\mu^3 \phi_N$ instead of $\mu^3 \phi_0$, the theory would still have an exact $\mathbb{Z}_{3^N}$ discrete global symmetry, as pointed out in \Ref{Dvali:2007hz, Dvali:2008tq}. If we now take $\phi_0$ to be light and integrate out all the other $\phi_k$, then we will obtain an EFT where the lowest dimension operator violating $\phi_0$'s continuous $U(1)$ is necessarily $\propto \phi_0^{\left(3^N\right)}$, which arises at tree level. This exponentially large operator dimension could easily violate the bound \Eq{eq:dimbound}, even accounting for the constraint on $N$ from the ``species bound'' \Eq{eq:sqrtN}. This reflects the observation in \Sec{sec:discrete} that arbitrary exact discrete symmetries can result in violation of the local rate bound. Note however that in this case there is a completion of the EFT, the full clockwork model with all the $\phi_k$, in which the SSC is restored. This occurs due to the exponentially large charges of some of the $\phi_k$ under the $\mathbb{Z}_{3^N}$ symmetry, which allows us to write low-dimension $\Uc$-violating operators.\,\footnote{\Ref{Dvali:2008tq} notes that the existence of these large charges also implies that this scenario is not immediately in tension with their arguments against large discrete symmetries.}

\subsection{Extradimensional localization}
\label{sec:flat}

\subsubsection{A flat interval}

Wavefunction localization in a compact extra-dimension with broken Poincar\'e invariance provides another mechanism generating near-exact global symmetries.

Let us consider a fifth dimension  compactified as an interval of length $L$, with coordinate $y\in[0,L]$. The five-dimensional reduced Planck mass $M_5$ is related to the 4D one as $M^3_5L=\mPlr^2$ and should satisfy $M_5 L\gtrsim 1$. The cutoff of the 5D theory is set by dimensional analysis to be $\Lambda_5=24^{1/3}\pi M_5\approx 10 M_5 $. The theory has Kaluza-Klein (KK) modes of mass $m^2_n \sim n^2\pi^2/L^2  $,

Branes are centered at the $y=0,L$ endpoints.  In  5D model-building it is commonplace to have brane-localized fields, which appear in the 5D Lagrangian in the form ${\cal L}_{\rm 5D}\supset \delta(y){\cal L}_{\rm brane}$ and similarly for the $y=L$ brane. Even though such terms are sufficient for phenomenological purposes, they are by no means the only Poincar\'e-violating terms allowed, a whole series of higher dimensional terms proportional to $(\partial_y)^n\delta(y)$ are allowed as well. These terms encode the finite width and shape of the brane, which are irrelevant at low energy but are relevant in our analysis because we tend to push the model in extreme corners. 

The fact that branes are not infinitely thin in the 5D EFT implies that there is no such thing as an infinitely localized brane Lagrangian. Hence brane fields should be understood as highly localized fields, with a tiny but finite wavefunction tail in the bulk. A similar conclusion is reached if one tries to infinitely localize a bulk field zero mode by sending its 5D mass to infinity (see expressions below). The 5D cutoff prevents to take such limit, so that again a brane field should be rather understood as a highly localized bulk field. Interestingly, if one ignores the 5D cutoff and send the bulk mass to infinity, inconsistencies appear because gravitons couple to the 5D mass. In the 4D effective theory, the coefficients of KK graviton-induced operators such as $(\partial^\mu \phi\,\partial_\mu \phi)^2$ tend to infinity, rendering the 4D EFT description invalid (see \cite{Fichet:2013ola}).

These considerations readily eliminate a paradox which would have happened if exactly localized brane fields were possible. If such exact localization was true, one may have a Lagrangian with a single species, say  complex scalars $\Phi_1$, $\Phi_2$, on each brane. Assume the scalars  are charged under a bulk Abelian gauge field. They interact with each other via the gauge field but cannot couple to each other directly. Hence in the 4D effective theory, only monomials in $|\Phi_1|^2$, $|\Phi_2|^2$  arise  from heavy Abelian KK modes exchange, while no operators such as $\Phi_1 \Phi_2 + \textrm{h.c.}$ appear. As a result the low-energy theory has two  exact global symmetries for $\Phi_1$ and $\Phi_2$ separately---while the gauge symmetry acts simultaneously on $\Phi_1$ and $\Phi_2$. 
Exact brane localization would therefore violate the SSC and more generally the quantum gravity lore that no exact global symmetries is allowed exist. This paradox is resolved for bulk fields, both as a result of the  zero mode profiles and of the existence of the $\Phi_{1,2}$ KK modes.

Let us thus focus on bulk fields. Consider a  bulk scalar $\Phi$ with a 5D mass, ${\cal L}_{5}\supset m^2_\Phi |\Phi|^2 $, which is allowed to be as high as the 5D cutoff, $m^2_\Phi \lesssim \Lambda^2_5$. 
 The KK modes of the scalar field have mass $m^2_n=n^2\pi^2/L^2 +m^2_{\Phi} $ and can thus be near the 5D cutoff.   The scalar can also have brane localized mass terms, and when masses on both branes are tuned to a value $\propto m_\Phi$ a massless mode is present in the spectrum, $\Phi=f_0(y)\phi^{(0)}(x)+\ldots$ The massless mode can be  localized towards either branes, with an exponential profile $f_0(y)\propto e^{-|m_{\Phi}| y}$.

Consider two scalar fields $\Phi_{1,2}$ with  5D masses $m_{1,2}$ interacting via bulk operators violating a global symmetry of the 5D Lagrangian. Let us localize the zero mode of $\Phi_1$ towards one brane and that of $\Phi_2$ towards the other. The 4D coupling between massless modes is then exponentially suppressed, with coefficient $c_4 \sim e^{-(|m_1|+|m_2|)L}$. Taking the masses as high as possible, the global symmetry-violating operator involving the zero modes is suppressed as  $c_4\sim e^{-\Lambda_5L}$.

We can already see that the limit of exact localization corresponds to sending $\Lambda_5L$ to infinity, which would imply $M^2_{\rm p}/M_5^2\rightarrow \infty$ and thus render the 5D cutoff arbitrarily low. Thus the limit of an exact global symmetry cannot be taken. This obstruction resembles the one implied by the WGC when trying to send a gauge coupling to zero.

Let us then consider the local rate bound, first from the omniscient viewpoint of the model builder. Global symmetry violation by thermal QFT processes receives $e^{-\Lambda_5L}$-suppressed contributions from zero mode interactions, while thermal  processes from KK modes are Boltzmann suppressed as $e^{-\Lambda_5/T}$. Depending on temperature, either the zero mode or the KK mode contributions dominate. For any temperature,  these contributions are always larger than the black hole symmetry violation rate, since the smallest black holes have to  be substantially heavier than $\Lambda_5$. Therefore the local rate bound is not violated when considering the full 5D theory. 

Let us then take the viewpoint of a low-energy observer carrying out an experiment at zero temperature. At energies  below the KK scale $\sim \pi/L$, the observer knows only about the scalar  zero modes and the coefficient $c_4$. Moreover the observer knows only about 4D gravity. What does the observer conclude when using the 4D SSC bound? Using the log-coefficient bound Eq.~\eqref{eq:logcbound} on  $c_4$, the observer concludes that the local rate bound is violated  when the \textit{a priori} unknown cutoff scale is taken above the value $\Lambda_{\rm SSC}\sim 6 M_5$.  Since $M_5<\mPl$, the observer can suspect that new symmetry-violating effects exist below the Planck scale. In fact, the SSC bound gives precisely the  mass scale at which the scalar KK modes responsible of the symmetry violation would actually appear.

Increasing the experiment's energy scale, the observer then discovers KK gravitons at $E\sim \pi/L$. Hence the observer may strongly suspect that they are seeing 5D gravity. However, for energies $\lesssim \Lambda_5$, a 4D EFT involving towers of KK states can still provide a controlled EFT description. From this 4D EFT perspective the SSC is satisfied as discussed above. On the other hand, one could also choose to write manifestly a 5D EFT to describe physics at scales above $\sim \pi/L$. A 5D description indicates the presence of 5D black holes which can be much lighter than $\mPl$, so that the thermal rate bound becomes stronger than in the 4D case and the log-coefficient bound Eq.~\eqref{eq:logcbound} is no longer satisfied. However, in order to maintain local Lorentz invariance in the 5D theory (which we require as a condition for the SSC, \Sec{se:swamp}) the $\Phi_{1,2}$ must be described as 5D bulk fields. This implies the existence of their scalar KK modes at a scale $\lesssim \Lambda_5$, \textit{i.e.} below the 5D black hole mass, so that they mediate violation of the global symmetry fast enough to satisfy the local rate bound. Hence, though the local rate bound takes a different form depending on whether a 4D or 5D description is used, it is satisfied in both cases. 

\subsubsection{A warped interval}
\label{sec:warped}

The case of a ``slice'' of AdS is similar to the flat background case in certain aspects. Namely, the exponential localization of massless modes and the new physics scale expected from a 4D observer evaluating the local rate bound are essentially the same as in the flat case. However there are also  qualitative differences between curved and flat space which deserve to be discussed.  

Let us consider a truncated   AdS background of length $L$ and curvature $k$, with coordinate $y\in[0,L]$.  Assuming substantial warping $e^{kL}\gg 1$, the five-dimensional reduced Planck mass is related to the 4D one as $M^3_5=\mPlr^2k$ and should satisfy $M_5\lesssim \mPlr$, \textit{i.e.} $k\lesssim \mPlr$. 

Like in the flat case we can consider a bulk scalar, whose 5D mass can be as high as the 5D cutoff. The mass of scalar KK modes is, however, much lower than in the flat case. Even for the largest bulk mass possible, the first scalar KK modes   have mass $ \sim \frac{\pi}{2} m_\Phi e^{-\pi k L}  $,  close to the KK scale. This is a striking difference with the flat case; while a low-energy observer may expect symmetry violation around $\Lambda_5$ by using the local rate bound, it turns out that the new symmetry-violating  physics always appears at a much lower scale. Therefore the observer's conclusion drawn from the local rate bound was very conservative. 

It is also interesting to understand what an observer actually sees when increasing the energy of the experiment.  A deep difference between EFT in AdS and in flat space appears there.  At four-momentum higher than $\tilde \Lambda_5 \equiv \Lambda_5 e^{-\pi k L} $, the KK mode decomposition becomes invalid because the 5D EFT breaks down near the IR brane, \textit{i.e.} gravity becomes strongly coupled \cite{Goldberger:2002cz}. Therefore we cannot formulate the EFT  in terms of KK variables, and we instead should use the 5D language. For an observer on the IR brane,  the Planckian completion of 5D gravity appears at the scale $\tilde \Lambda_5$. This scale can be higher or nearly equal to the first KK mode mass. Hence the observer sees a violation of the emergent symmetry either because of the scalar KK modes or directly because of the Planckian  completion. 
Instead, for an observer on the UV brane, the EFT  corresponds to the holographic effective action used for the AdS/CFT 
correspondence. Considerations about the global symmetry are nontrivial in this case because the IR brane degrees of freedom are not manifest in the holographic action. They are rather present via nonlocal effects, which is beyond the scope of application of the SSC. 
Moreover at momentum above $\tilde \Lambda$, the IR brane tends to vanish from the UV correlators \cite{Goldberger:2002cz,Fichet:2019hkg}. Similarly, at finite temperature, a Hawking-Page-like transition towards an AdS-Schwarzschild metric is expected, which removes the IR brane \cite{Creminelli:2001th}.  An analysis of these aspects is beyond the scope of the present work. 

\section{Conclusions}
\label{se:conc}

How exact can a global symmetry be while remaining consistent with quantum gravity? A path to answer this fundamental question goes through the study of black holes, which in a sense are a window into Planckian physics. However, the general quantum theory of black holes is unknown, which limits our possibilities and renders the problem highly challenging. 

In this work we point out that one could approach this by  considering the theory at finite (sub-Planckian) temperature, in which black holes exist with some small population and can be safely described semiclassically. This in turn can be used to build an argument to quantitatively constrain exactness of global symmetries in the QFT, using the hypothesis that  local symmetry-violating effects should be at least as important as Boltzmann-suppressed black hole effects.

To give a more concrete sense to these arguments,  we consider a thought experiment involving a static universe at finite temperature, making sure to account for instability from black hole nucleation. The thermal thought experiment provides a consistent setup in which symmetry violation is induced both by black holes and by local QFT operators.

We study in detail the rate of black hole formation from 2-body scattering as well as from thermal fluctuations, and argue that a thermal density of small black holes forms well before the instability of the space sets in. Particle capture by this thermal density of black holes is a well-understood low-energy process that causes violation of global symmetry in the thermal bath. Our calculation thus provides an irreducible rate of symmetry violation in a thermal system.

Building on these results we introduce the ``local rate bound'', the statement that  local EFT symmetry-violating processes are at least as large as the black hole ones 
for any temperature allowed by the thought experiment.  We then formulate a ``Swampland Symmetry Conjecture'' (Sec.~\ref{se:swamp}) applying to effective QFTs with any sub-Planckian validity cutoff.

We study the interplay of the SSC with QFT models featuring emergent symmetries.  Models  with accidental symmetries from gauged $U(1)$'s are an especially interesting playground since these are constrained by the Weak Gravity Conjecture. It turns out that appropriate forms of the WGC can ensure that the SSC is satisfied by mandating the existence of enough charged states to break any accidental symmetries at the necessary level.

We study the consistency of our conjecture in QFT models with low-energy emergent symmetries such as the Froggatt-Nielsen mechanism, extradimensional localization and ``clockwork." Each of these mechanisms can give exponentially exact global symmetries in a certain limit. However in each case a low-energy observer knowing only about the near-exact symmetry can use the local rate bound to predict an upper bound $\Lambda_\mathrm{SSC}$ on the cutoff scale for which the theory satisfies the SSC. In almost all cases this $\Lambda_\mathrm{SSC}$ is indeed higher than the actual cutoff scale of the theory (the only exception we found occurs in a case with an exponentially large discrete symmetry, \Sec{sec:clock}).

The Swampland Symmetry Conjecture is an attempt to generalize the various examples we have analyzed to a constraint on EFTs in the spirit of the swampland program. As it is stated here, it does not directly constrain EFTs defined below the Planck scale, since it just requires that some completion of the EFT satisfy the local rate bound \Eq{eq:TRB}. However, in the various QFT models of emergent symmetry which we have discussed, the local rate bound is often guaranteed to hold even in the most low-energy EFT descriptions, so that the SSC is manifestly satisfied. In particular, there are multiple cases in which the local rate bound can be parametrically saturated, but not violated.

We may compare this situation to how the Weak Gravity Conjecture constrains the space of gauge EFTs consistent with gravity. The most minimal version of the WGC can be satisfied by introducing a superextremal particle with large charge and mass near the Planck scale, so that EFTs defined below $\mPl$ are technically unconstrained. However in most examples some much stronger statement such as the Lattice WGC holds, which causes the WGC-satisfying particles to appear with the lowest possible charge, \textit{i.e.} at masses below $g \mPl$ so that they appear within low-energy EFT. Similarly, if the SSC is true so that the local rate bound is always satisfied within the maximal sub-Planckian EFT, we may expect that it is usually satisfied in lower-energy EFTs as well.

Judging by the history of various swampland conjectures, one may anticipate that the statement of the SSC could potentially be refined in various ways---perhaps additional assumptions are necessary to guarantee it, or more input from quantum gravity could give a stronger constraint. Studying more examples of emergent global symmetry could help achieve this. In this work we mainly focused on QFT models, with the only quantum gravity input being other swampland conjectures such as the WGC and the species bound \Eq{eq:sqrtN}. Studies of string theory models producing approximate global symmetries could provide new examples (or counterexamples!) that would help determine the status of the SSC.  

The quantitative results on symmetry violation that we have discussed in this work provide a bridge between quantum gravity and emergent symmetry. 
  It is clear however that there is much to be understood about the status of  symmetries within quantum gravity and its landscape of EFTs.

 \section*{Acknowledgements}

  SF is supported by the S\~ao Paulo Research Foundation (FAPESP) under grants \#2011/11973, \#2014/21477-2 and \#2018/11721-4. PS is supported by the DuBridge Fellowship of the Walter Burke Institute for Theoretical Physics. This material is based upon work supported by the U.S. Department of Energy, Office of Science, Office of High Energy Physics, under Award Number DE-SC0011632.

\appendix

\section{Non-relativistic $n$-body process }

\label{app:NR_EFT} 

The generic thermal rate is given by
\begin{align}
\Gamma= & \kappa_i \kappa_f \prodsym_{i=1}^n  \frac{d^3{\bf p}_i}{(2\pi)^3 2E_i}
\prodsym_{f=1}^k  \frac{d^3{\bf p}_f}{(2\pi)^3 2E_f}\, f_i \, |{\cal M}|^2 
 \,(2 \pi)^4\,\,\delta^{(4)}\left(\sum^n_{i=1} { p}_i-\sum^k_{f=1} { p}_f\right) \,.\label{eq:rate_gen}
\end{align}
In the Boltzmann equation for mean number densities, the relevant statistics $f_i$ are
\be
f_i=\frac{g_*}{e^{E_i/T}\pm 1}
\ee
for fermions/bosons. 

The non-relativistic $\{n\}\rightarrow\{k\}$ process induced by an operator of the kind of Eq.~\eqref{eq:OGv} with no derivatives  gives rise to the  rate 
\begin{align}
\Gamma= \kappa_i \kappa_f e^{- nm\beta} & \frac{|{\cal M}|^2}{(4 \pi^3 m)^{n+k}} \prodsym_{i=1}^n d^3 {\bf p}_i
\prodsym_{f=1}^k d^3 {\bf p}_f e^{-\beta \sum^n_{i=1} |{\bf p}|^2_i/2m} \,\,\, \times \nonumber  \\
 &(2 \pi)^4\,\,\delta^{(3)}\left(\sum^n_{i=1} {\bf p}_i-\sum^k_{f=1} {\bf p}_f\right) \delta\left(
\frac{\sum^n_{i=1}|{\bf p}|^2_i}{2m}-\frac{\sum^k_{f=1}|{\bf p}|^2_f}{2m}+O(k-n)m
\right)
\end{align}
with $|{\cal M}|^2=1/\Lambda^{2n_{\not G}-8}$.

Using the $SO(n)$ symmetry in the initial states, one can easily integrate over the $\delta^{(3)}$ Dirac, which gives
\begin{align}
\Gamma= \kappa_i \kappa_f e^{- nm\beta} & \frac{|{\cal M}|^2}{(4 \pi^3 m)^{n+k}} \prodsym_{i=1}^{n-1} d^3 {\bf p}_i
\prodsym_{f=1}^k d^3 {\bf p}_f e^{-\beta \sum^n_{i=1} |{\bf p}|^2_i/2m} \,\,\, \times \nonumber  \\
 & \frac{(2 \pi)^4}{n^{3/2}} \delta\left(
\frac{\sum^{n-1}_{i=1}|{\bf p}|^2_i}{2m}-\frac{\sum^{k-1}_{f=1}|{\bf p}|^2_f}{2m}
+O(k-n)m\right) \,.
\end{align}
The remaining integral in the initial momenta is closely related to the surface of a $(3n-4)-$sphere  with radius 
\be
R^2= \sum^{k-1}_{f=1}|{\bf p}|^2_f \,.
\ee
We obtain
\begin{align}
\Gamma= \kappa_i \kappa_f e^{- nm\beta}  \frac{|{\cal M}|^2}{(4 \pi^3 m)^{n+k}} 
 \prodsym_{f=1}^k d^3 {\bf p}_f 
 e^{-\beta \sum^k_{f=1} |{\bf p}|^2_f/2m} \,\frac{m}{R} 
 R^{3n-4} S_{3n-4} 
\end{align}
where $S_a$ is  the surface of the unit $(a-1)-$sphere
\be
S_{a} = \frac{ a \pi^{a/2}}{\Gamma(a/2+1)}\,.
\ee
Note the radius depends on the $k-1$ final momenta. The integral over the  $k-$th momentum is Gaussian and gives
\begin{align}
\Gamma= \kappa_i \kappa_f e^{- nm\beta}  \frac{|{\cal M}|^2}{(4 \pi^3 m)^{n+k}} 
\frac{(2 \pi)^4}{n^{3/2}} (2\pi mT)^{3/2}
\prodsym_{f=1}^{k-1} d^3 {\bf p}_f & e^{-\beta R^2/2m} \,\frac{m}{R}R^{3n-4} S_{3n-4} \,.
\end{align}
The remaining integral is again evaluated in spherical coordinates, where the angle integrals give the surface of a $(3k-4)-$sphere $S_{3k-3}$, 
\begin{align}
\Gamma= \kappa_i \kappa_f e^{- nm\beta}  \frac{|{\cal M}|^2}{(4 \pi^3 m)^{n+k}} 
\frac{(2 \pi)^4}{n^{3/2}} (2\pi mT)^{3/2} m
\int dR & e^{-\beta R^2/2m} \,R^{3(n+k)-9} S_{3n-4} S_{3k-3} \,.
\end{align}
We can  rewrite this expression introducing $\beta$ derivatives
\begin{align}
\Gamma= \kappa_i \kappa_f e^{- nm\beta}  \frac{|{\cal M}|^2}{(4 \pi^3 m)^{n+k}} 
\frac{(2 \pi)^4}{n^{3/2}} (2\pi mT)^{3/2} m  S_{3n-4} S_{3k-3} \,s (- 2m \, \partial_\beta)^{3(n+k)/2-9/2}
\int dR & e^{-\beta R^2/2m} \,. 
\end{align}
The remaining integral is a trivial Gaussian, and the subsequent derivatives can be obtained from the identity
\be
(\partial_\beta)^n \beta^{-1/2} = \frac{2n!}{(-4^n)n!}\beta^{-1/2-n}\,.
\ee
Combining all the pieces gives the final rate Eq.~\eqref{eq:rate_EFT_NR}.

\bibliographystyle{JHEP} 

\bibliography{biblio}

\end{document}